\documentclass[letterpaper,12pt]{JHEP3}
\usepackage{amsmath}
\usepackage{amssymb}
\usepackage{bbm}
\usepackage{subfigure}
\usepackage{epsfig}
\usepackage{yfonts}
\newcommand{\labell}[1]{\label{#1}}

\newcommand{\be}{\begin{equation}}
\newcommand{\ee}{\end{equation}}
\newcommand{\bea}{\begin{eqnarray}}
\newcommand{\eea}{\end{eqnarray}}
\newcommand{\ba}{\begin{eqnarray}}
\newcommand{\ea}{\end{eqnarray}}

\newcommand{\beq}{\begin{equation}}
\newcommand{\eeq}{\end{equation}}
\newcommand{\beqa}{\begin{eqnarray}}
\newcommand{\eeqa}{\end{eqnarray}}
\newcommand{\beqar}{\begin{eqnarray*}}
\newcommand{\eeqar}{\end{eqnarray*}}

\newcommand{\reef}[1]{(\ref{#1})}

\newcommand{\eg}{{\it e.g.,}\ }
\newcommand{\ie}{{\it i.e.,}\ }

\newcommand{\mt}[1]{\textrm{\tiny #1}}

\newcommand{\X}{\mathcal{X}}

\newcommand{\C}{\mathcal{C}}

\newcommand{\tL}{\tilde{L}}

\newcommand{\la}{\lambda}

\newcommand{\lp}{\ell_{\mt P}}

\newcommand{\fin}{f_\infty}

\newcommand{\ct}{C_{T}} %{C_\mt{T}}

\newcommand{\del}{\partial}
\newcommand{\ads}{a_d^*}
\newcommand{\tc}{c} %{{\tilde c}}

\def\ov{\over}
\def\dr{\dot{r}}
\def\dtr{\dot{\tilde{r}}}
\def\l{\ell}
\def\ri{\right}

\def\({\left(}
\def\){\right)}
\def\[{\left[}
\def\]{\right]}

\newcommand{\dphi}{{\delta\phi}}
\newcommand{\dtphi}{\widetilde{\delta\phi}}

\preprint{arXiv:1202.2068 [hep-th]}

\title{Comments on Holographic Entanglement Entropy and RG Flows}

\author{Robert C. Myers$^a$ and Ajay Singh$^{a,b}$ \\
$^a$ {\it Perimeter Institute for Theoretical Physics, Waterloo,
Ontario N2L 2Y5, Canada}\\
$^b$ {\it Department of Physics \& Astronomy and Guelph-Waterloo Physics Institute,}\\
{\it \ \ University of Waterloo, Waterloo, Ontario N2L 3G1, Canada} \\
}
\abstract{Using holographic entanglement entropy for strip geometry, we construct
a candidate for a c-function in arbitrary dimensions. For holographic theories dual
to Einstein gravity, this c-function is shown to decrease monotonically along RG flows.
A sufficient condition required for this monotonic flow is that the stress tensor of
the matter fields driving the holographic RG flow must satisfy the null energy condition
over the holographic surface used to calculate the entanglement entropy.
In the case where the bulk theory is described by Gauss-Bonnet gravity, the latter condition alone
is not sufficient to establish the monotonic flow of the c-function. We also observe
that for certain holographic RG flows, the entanglement entropy undergoes a `phase transition'
as the size of the system grows and as a result, evolution of the c-function may exhibit
a discontinuous drop.
}

\begin{document}

%%%%%%%%%%%%%%%%%%%%%%%%%%%%%%%%%%%%%%%%%%%%%%%%%%%%%%%%%%%%%%%%%%%
\section{Introduction} \label{intro}

Zamolodchikov \cite{zamo} showed that renormalization group (RG) flows
of two-dimensional quantum field theories were governed by a remarkable
underlying structure. One important feature was that there exists a
positive definite function $c_2$, which decreases monotonically along
the RG flows. At the fixed points of the RG flow, this function is
stationary and coincides with the central charge $c$ of the conformal
field theory (CFT) describing the fixed point. A direct consequence for
any RG flow connecting two such fixed points is then that
 \be
\big[\,c\,\big]_\mt{UV}\, \geq\, \big[\,c\,\big]_\mt{IR}\,.
 \labell{introx1}
 \ee
More recently, Casini and Huerta \cite{casini2} developed an elegant
reformulation of Zamolodchikov's c-theorem in terms of entanglement
entropy in two dimensions. In their construction, the c-function was
defined as
 \be
c_2\,=\,3\, \l\, {d S(\l)\ov d\l}\,, \labell{introx2}
 \ee
where $S(\l)$ denotes the entanglement entropy for an interval of
length $\ell$. Then it follows that $dc_2/d\l \leq 0$ from the strong
subadditivity property of entanglement entropy, as well as the Lorentz
symmetry and unitarity of the underlying quantum field theory (QFT).
Therefore, as the QFT is probed at longer distance scales, \ie one
increases $\l$, this c-function \reef{introx2} decreases monotonically.
Further, for a two-dimensional CFT, the entanglement entropy is given
by \cite{cardy0,cardyCFT}
 \be
S_\mt{CFT}=\frac{c}{3} \,\log(\,\ell/ \delta) + c'\,,
 \labell{twod}
 \ee
where $c$ is the central charge, $\delta$ is a short-distance regulator
and $c'$ is a non-universal constant (independent of $\ell$). Hence
$c_2=c$ at RG fixed points.

As a generalization of the two-dimensional c-theorem, Cardy
\cite{cardy} conjectured that the central charge associated with A-type
trace anomaly -- see eq.~\reef{trace} -- should decrease monotonically
along RG flows for QFT's in any even number of dimensions. Of course,
in two dimensions, this proposal coincides precisely with
Zamolodchikov's result \reef{introx1} since $c=12\,A$. Cardy's
conjecture was extensively studied in $d=4$ and a great deal of support
was found with nontrivial examples, including perturbative fixed points
\cite{jack} and supersymmetric gauge theories
\cite{anselmi1,anselmi2,ken}.\footnote{Note that a flaw was recently
found \cite{yuji2} in a proposed counter-example \cite{yuji} to Cardy's
conjecture.} Recently, a remarkable new proof of this c-theorem was
presented for any four-dimensional RG flow connecting two conformal
fixed points \cite{zohar}. This result draws on earlier work involving
the spontaneous breaking of conformal symmetry \cite{spon} and bounds
on couplings in effective actions \cite{fast}. It remains to determine,
however, how much more of the structure of two-dimensional RG flows
carries over to higher dimensions.\footnote{A related question which
has seen active discussion in the recent literature is whether or not
there exist interesting QFT's in higher dimensions which exhibit scale
invariance but not conformal invariance \cite{scale,jeff}. Of course,
in two dimensions, it is proven that scale invariant QFT's are also
conformally invariant \cite{joep}.}

As we will review below, support for Cardy's generalized c-theorem was
also established using the AdS/CFT correspondence
\cite{gubser,flow2,cthem}. One of the advantages of the investigating
RG flows in a such holographic framework is that the results are
readily extended to arbitrary dimensions . In particular then, the
analysis of holographic RG flows identified a certain quantity
satisfying an inequality analogous to eq.~\reef{introx1} for {\it any
dimension}, that is, for both odd and even numbers of spacetime
dimensions. Since the trace anomaly is only nonvanishing for even $d$,
a new interpretation was required for odd $d$. Ref.~\cite{cthem}
identified the relevant quantity as the coefficient of a universal
contribution to the entanglement entropy for a particular geometry in
both odd and even $d$. These holographic results then motivated a
generalized conjecture for a c-theorem for RG flows of odd- and
even-dimensional QFT's. For even $d$, this new central charge was shown
to precisely match the coefficient of the A-type trace anomaly
\cite{cthem} and so this conjecture coincides with Cardy's proposal.
For odd $d$, it was shown that this effective charge could also be
identified by evaluating the partition function on a $d$-dimensional
sphere \cite{casini9} and so the conjecture is connected to the newly
proposed F-theorem \cite{fthem}.

The above developments motivated the present paper which examines the
the connections between entanglement entropy and RG flows in a
holographic framework. Earlier work in this direction can be found in
\cite{ent2,calc,albash}. Here, we make a simple generalization of the
c-function in eq.~\reef{introx2} to higher dimensions and then use a
holographic framework to examine its behaviour in RG flows. We are able
to show that subject to specific conditions, the flow of the c-function
is monotonic for boundary theories dual to Einstein gravity. In
examining specific flow geometries, we also find that the entanglement
entropy undergoes a `first order phase transition' as the size of the
entangling geometry passes through a critical value. That is, in our
holographic calculation, there are competing saddle points and the
dominant contribution shifts from one saddle point to another at the
critical size.

An overview of the paper is as follows: In section \ref{review}, we
review the standard derivation of holographic c-theorems with both
Einstein gravity and Gauss-Bonnet gravity in the bulk. We stress that
in either case, the monotonic flow of the c-function requires that the
matter fields driving the holographic RG flow must satisfy the null
energy condition. In section \ref{ent}, we discuss the holographic
entanglement entropy for the `strip' or `slab' geometry and construct a
c-function which naturally generalizes eq.~\reef{introx2} to higher
dimensions. In section \ref{one}, we show that for an arbitrary RG flow
solution in Einstein gravity, this c-functions decreases monotonically
if the bulk matter fields satisfy the null-energy condition. Section
\ref{examp1} considers explicit examples of holographic RG flows and
demonstrates that in certain cases, the entanglement entropy undergoes
a `phase transition.' As a result, the c-function exhibits a
discontinuous drop along these RG flows. In section \ref{three}, we
examine holographic RG flows with Gauss-Bonnet gravity and there, we
find that the null-energy condition is insufficient to constrain the
flow of our c-function to be monotonic. We conclude with a brief
discussion of our results and future directions in section
\ref{discuss}. Appendix \ref{two} presents certain technical details
related to the discussion in section \ref{one}. In the appendix
\ref{gravity}, we discuss holographic RG flow solutions in Gauss-Bonnet
gravity. Finally, appendix \ref{bulk} describes the construction of a
bulk theory for which the holographic flow geometries examined in
section \ref{examp1} would be solutions of the equations of motion.

While we were in the final stages of preparing this paper, we learned
of a similar study of entanglement entropy and holographic RG flows
appearing in \cite{friends}.

%%%%%%%%%%%%%%%%%%%%%%%%%%%%%%%%%%%%%%%%%%%%%%%%%%%%%%%%%%%%%%%%%%%%
\section{Review of holographic c-theorems} \label{review}

Here we begin with a review of the holographic c-theorem as originally
studied by \cite{gubser,flow2} for Einstein gravity. These references
begin by constructing a holographic description of RG flows. The
simplest case to consider is ($d$+1)-dimensional Einstein gravity
coupled to a scalar field:
 \be
I=\frac{1}{2\lp^{d-1}}\int d^{d+1} x \, \sqrt{-g} \left(R-\frac{1}{2}
\left(\del \phi\right)^2-V(\phi)\right)
 \labell{action}
 \ee
We assume that the potential $V(\phi)$ has various critical points
where the potential energy is negative, \ie
  \be
V(\phi_i)=-\frac{d(d-1)}{L^2}\alpha_i^2 \qquad {\rm where}\ \ \
\left.\frac{\delta V}{\delta \phi}\right|_{\phi=\phi_i}=0\,.
 \labell{crit2}
  \ee
Here $L$ is some convenient scale while the dimensionless parameters
$\alpha_i$ distinguish the different fixed points. At these points, the
gravity vacuum is simply AdS$_{d+1}$ with the curvature scale given by
$\tL^2=L^2/\alpha_i^2$.

Now in the context of the AdS/CFT correspondence, the bulk scalar above
is dual to some operator $\mathcal{O}$ and the fixed points
\reef{crit2} of the scalar potential represent the critical points of
the boundary theory. In particular then, with an appropriate choice of
the bulk potential, $\mathcal{O}$ will be a relevant operator for a
certain fixed point and so an RG flow will be triggered by perturbing
the corresponding critical theory by this operator in the UV. Of
course, the holographic description of this RG flow is that the scalar
field acquires an nontrivial radial profile which connects two of the
critical points in eq.~\reef{crit2}. The bulk geometry for this
solution can be described with a metric of the following form
\cite{gubser,flow2}
 \be
ds^2=e^{2A(r)}\,\eta_{ij}\,dx^i dx^j+dr^2\,.
 \labell{metric0}
 \ee
Here, the radial evolution of the geometry is entirely encoded in the
conformal factor $A(r)$. At a fixed point where the geometry is
AdS$_{d+1}$, the conformal factor is simply $A(r)=r/\tL$ where again
$\tL$ is the AdS curvature scale. Implicitly, we will assume that
asymptotic UV boundary is at $r\to \infty$ while the IR part of the
solution corresponds to $r\to -\infty$. Hence for an RG flow between
two fixed points as described above, the metric \reef{metric0}
approaches that of AdS$_{d+1}$ in both of these limits.

Now following \cite{gubser,flow2}, we define:
\be a_d(r)\equiv\frac{\pi^{d/2}}{\Gamma(d/2)\,\left(\lp
A'(r)\right)^{d-1}}\,, \labell{def0} \ee
where `prime' denotes a derivative with respect to $r$. Then for
general solutions of the form \reef{metric0}, one finds
\bea a_d'(r)&=&-\frac{(d-1)\pi^{d/2}}{\Gamma(d/2)\,\lp^{d-1} A'(r)^d}\,
A''(r)
\labell{magic0}\\
&=& -\frac{\pi^{d/2}}{\Gamma(d/2)\,\lp^{d-1}
A'(r)^d}\,\left(T^t{}_t-T^r{}_r\right) \ge0\,.\nonumber
 \eea
Above in the second equality, Einstein's equations were used to
eliminate $A''(r)$ in favour of components of the stress
tensor.\footnote{Note that for the scalar field theory in
eq.~\reef{action}, we have $T^t{}_t-T^r{}_r=-(\phi')^2/2\le0$.} The
final inequality assumes that the matter fields obey the null energy
condition \cite{HE}. Now given the usual connection between $r$ and
energy scale in the CFT, eq.~\reef{magic0} indicates that $a(r)$ is
always increasing as we move from low energies to higher energy scales.
Further, if the flow function \reef{def0} is evaluated for an AdS
background, one finds a constant:
 \be
\ads=a_d(r)\big|_\mt{AdS} =\frac{\pi^{d/2}}{\Gamma(d/2)}\,\frac{\tilde
L^{d-1}}{\lp^{d-1}}\,. \labell{acharge}
 \ee
Hence if we compare this constant for the UV and IR fixed points of the
holographic RG flow, we find the holographic c-theorem:
 \be
\big[\,\ads\,\big]_\mt{UV}\ \ge\ \big[\,\ads\,\big]_\mt{IR}
 \labell{theorem0}
 \ee

To make closer contact with the dual CFT, we recall the trace anomaly
\cite{traca,deser},
 \be
\langle\,T^i{}_i\,\rangle = \sum_n B_n\,I_n -2\,(-)^{d/2}A\, E_d\,,
 \labell{trace}
 \ee
which defines the central charges for a CFT in an even number of
spacetime dimensions. Each term on the right-hand side is a Weyl
invariant constructed from the background geometry. In particular,
$E_d$ is the Euler density in $d$ dimensions while the $I_n$ are
naturally written in terms of the Weyl tensor (as well as its covariant
derivatives), \eg see \cite{fefferman}. Note that in eq.~\reef{trace},
we have ignored the possible appearance of a conformally invariant but
also scheme-dependent total derivative.

A holographic description of the trace anomaly was developed
\cite{sken} and can be applied to the AdS$_{d+1}$ stationary points in
the present case (for even $d$). These calculations show that $\ads$,
the value of the flow function at the fixed points, precisely matches
the A-type central charge in eq.~\reef{trace}, \ie $\ads=A$ for even
$d$ \cite{cthem}. Hence with the assumption that the matter fields obey
the null energy condition, the holographic CFT's dual to Einstein
gravity satisfy Cardy's conjecture of a c-theorem for quantum field
theories in higher dimensions \cite{cardy}. Of course, one must add the
caveat that for these holographic CFT's, \ie those dual to Einstein
gravity, all of the central charges in eq.~\reef{trace} are equal to
one another \cite{sken}. Hence the holographic models \reef{action}
considered above can not distinguish between the behaviour of $A$ and
$B_n$ in RG flows.

It has long been known that to construct a holographic model where the
various central charges are distinct from one another, the gravity
action must include higher curvature interactions \cite{highc}. In
part, this motivated the recent holographic studies of Gauss-Bonnet
(GB) gravity \cite{lovel} --- for example, see \cite{EtasGB}. In
section \ref{three}, we will extend our discussion of holographic RG
flows to GB gravity with the following action
 \be
I = \frac{1}{2\lp^{d-1}} \int d^{d+1} x \sqrt{-g} \left( R +
\frac{\lambda L^2}{(d-2)(d-3)} \X_4 -\frac12(\partial\phi)^2-V(\phi)
\right) ~,
 \labell{GBAction}
 \ee
where
 \be
\X_4=R_{abcd}R^{abcd}-4R_{ab}R^{ab}+R^2\,.
 \labell{GBterm}
 \ee
As before, we again assume the scalar potential has various stationary
points as in eq.~\reef{crit2}, where the energy density is negative.
Note that for convenience, we are using the same canonical scale $L$
which appears for the critical points in eq.~\reef{crit2} in the
coefficient of the curvature-squared interaction in
eq.~\reef{GBAction}. Hence the strength of this GB term is controlled
by the dimensionless coupling constant, $\lambda$. We write the
curvature scale $\tL$ of the AdS vacuum as $\tL^2=L^2/\fin$ where the
constant $\fin$ satisfies \cite{cthem}
  \be
 \alpha_i^2-\fin+\lambda \fin^2=0\,.
\labell{curvature}
  \ee
In general, eq.~\reef{curvature} has two solutions but we only consider
the smallest positive root
  \be
\fin=\frac{1-\sqrt{1-4\lambda\,\alpha_i^2}}{2\lambda}\,,
\labell{physical}
  \ee
with which, in the limit $\la\to0$, we recover $\fin=\alpha_i^2$ and
$\tL^2=L^2/\alpha_i^2$, as discussed above for Einstein gravity. One
would find that graviton fluctuations about the AdS solution
corresponding to the second root are ghosts \cite{GBghost,old1} and
hence the boundary CFT would not be unitary. The theory \reef{GBAction}
is further constrained by demanding that the dual boundary theory
respects micro-causality or alternatively, that it does not produce
negative energy fluxes \cite{EtasGB,cc}.

For our present purposes, the most important feature of GB gravity
\reef{GBAction} is that the dual boundary theory will have two distinct
central charges. To facilitate our discussion for arbitrary $d\ge4$, we
would like to define two central charges that appear in any CFT for any
$d$ -- including odd $d$ -- and hence for our pursposes, the trace
anomaly is not a useful definition of the central charges. Following
\cite{cc,cthem}, we consider:
 \bea
\ct&=& \frac{\pi^{d/2}}{\Gamma(d/2)} \(\frac{\tL}{\lp}\)^{d-1} \left[ 1
- 2 \lambda \fin \right]\,,
 \labell{effectc}\\
\ads &=&\frac{\pi^{d/2}}{\Gamma(d/2)} \(\frac{\tL}{\lp}\)^{d-1} \left[
1 - 2\frac{d-1}{d-3} \lambda \fin \right]\,. \labell{effecta}
 \eea
The first charge $\ct$ is that controlling the leading singularity of
the two-point function of the stress tensor.\footnote{Here, as in
\cite{renyi}, we have normalized $\ct$ so that in the limit $\la\to0$,
$\ct=\ads$. This choice is slightly different from that originally
presented in \cite{cc}, \ie
$\ct|_{\cite{cc}}=\frac{d+1}{d-1}\frac{\Gamma(d+1)}{\pi^d}\,\ct|_\mt{here}$.}
The second central charge $\ads$ can be determined by calculating the
entanglement entropy across a spherical entangling surface
\cite{cthem}. Using the results of \cite{adam}, it was further shown
\cite{cthem} that $\ads$ is the central charge appearing in the A-type
trace anomaly in even dimensions, \ie $\ads=A$ in eq.~\reef{trace}. In
terms of these central charges, the micro-causality constraints,
referred to previously, are conveniently written as \cite{cc}
 \be
\frac{d(d-3)}{d(d-2)-2} \leq \frac{\ct}{\ads} \leq \frac{d}{2} \,.
 \labell{limitsca}
 \ee

Now assuming the existence of bulk solutions describing holographic RG
flows for the GB theory \reef{GBAction},\footnote{Appendix
\ref{gravity} includes a discussion of one approach to constructing
such solutions.} we can establish a holographic c-theorem following the
analysis of \cite{cthem}. We begin by constructing two flow functions
\cite{cthem}:
 \bea
\widehat{C}_T(r)&\equiv&\frac{\pi^{d/2}}{\Gamma(d/2)}\,\frac1{(
\lp\,A'(r))^{d-1}}\, \Big(1-{2}\lambda\,
L^2\,A'(r)^2\Big)\,, \labell{cfun}\\
a_d(r)&\equiv&\frac{\pi^{d/2}}{\Gamma(d/2)}\,\frac1{(
\lp\,A'(r))^{d-1}}\, \left(1-2\frac{d-1}{d-3} \lambda\, L^2A'(r)^2
\right)\,. \labell{afun}
 \eea
These expressions were chosen as the simplest extensions of
eq.~\reef{def0} which yield the two central charges above at the fixed
points, \ie $a_d(r)|_{AdS}=\ads$ and $\widehat{C}_T(r)|_{AdS}=\ct$
--- recall that $A(r)=r/\tL$ for the AdS vacua. Now let us examine the
radial evolution of $a_d(r)$ in a holographic RG flow:
 \bea
a_d'(r)&=&-\frac{(d-1)\,\pi^{d/2}}{\Gamma\left(d/2\right)\lp^{\,d-1}
A'(r)^d}\, A''(r)\, \Big(1-{2}\lambda\, L^2\,A'(r)^2  \Big)
\labell{magic}\\
&=& -\frac{\pi^{d/2}}{\Gamma\left(d/2\right)\lp^{\,d-1}
A'(r)^d}\,\Big(T^t{}_t-T^r{}_r\Big) \ge0\,.\nonumber
 \eea
Here, the equations of motion for GB gravity (see eq.~\reef{null}) have
been used to trade the expression in the first line for the components
of the stress tensor appearing in the second line. As before with
Einstein gravity, we assume the null energy condition applies for the
matter fields for the final inequality to hold. In eq.~\reef{GBAction},
the matter contribution is still a conventional scalar field action and
so just as before $T_t{}^t -T_r{}^r =-(\phi')^2/2 \le0$. With this
assumption, it then follows\footnote{We note that some additional
arguments are needed to ensure that there are no problems with
$A'(r)<0$ for odd $d$ \cite{cthem}.} that $a_d(r)$ evolves
monotonically along the holographic RG flows and we can conclude that
the central charge $\ads$ is always larger at the UV fixed point than
at the IR fixed point. Hence we recover precisely the same holographic
c-theorem found previously with Einstein gravity, namely,
 \be
\big[\,\ads\big]_\mt{UV}\ \ge\ \big[\,\ads\big]_\mt{IR}
 \labell{beta3}
 \ee

One can also consider the behaviour of $\widehat{C}_T$ along RG flows
 \be
\widehat{C}_T{}'(r)=-\frac{(d-1)\,\pi^{d/2}}{\Gamma\left(d/2\right)\lp^{\,d-1}
A'(r)^d}\, A''(r)\, \Big(1-{2}\frac{d-3}{d-1}\,\lambda\, L^2\,A'(r)^2
\Big)
 \labell{nomagic}
 \ee
but there is no clear way to establish that $\widehat{C}_T{}'(r)$ has a
definite sign. Hence this holographic model \reef{GBAction} seems to
single out $\ads$ as the central charge which satisfies a c-theorem.
This result has also been extended to holographic models with more
complex gravitational theories in the bulk:\footnote{Similar results
were also found to apply in the context of cosmological solutions
\cite{aninda9}.} quasi-topological gravity \cite{cthem}, general
Lovelock theories \cite{jtl,miguel}, higher curvature theories with
cubic interactions constructed with the Weyl tensor \cite{cthem} and
$f(R)$ gravity \cite{jtl}. The result is also established for
holographic models where the RG flow is induced by a double-trace
deformation of the boundary CFT \cite{double}. Given the relation
$\ads=A$ in even dimensions, these holographic results support Cardy's
proposal \cite{cardy} that the central charge $A$ (rather than any
other central charge) evolves monotonically along RG flows. However, it
is even more interesting that these results suggest that a similar
behaviour also occurs for the central charge $\ads$ in odd dimensions.
Further while the original field theory definition of $\ads$ involved a
calculation of entanglement entropy \cite{cthem}, it was shown that the
same charge can also be identified by evaluating the partition function
on $S^d$ \cite{casini9}. Hence the exciting new field theoretic results
of \cite{fthem} provide further evidence for the same c-theorem in odd
dimensions.

In any event, a key requirement for the holographic c-theorem to hold
for Einstein gravity \reef{theorem0} or for GB gravity \reef{beta3} is
that the matter fields obey the null energy condition. Of course, this
holds when these gravitational theories are coupled to a simple scalar
field, as in eqs.~\reef{action} and \reef{GBAction}, this constraint is
trivially satisfied. However, phrasing the constraint in terms of the
null energy condition allows for more general scenarios for the matter
fields driving the holographic RG flow. We should add that the same
constraint also ensures the holographic c-theorem holds for all of the
extensions of the bulk gravity theory mentioned above. We might mention
that violations of the null energy condition quite generally lead to
instabilities \cite{null} and so it is a natural constraint to define a
reasonable holographic model. In the following, we will also see that
the same constraint can be related to the monotonic flow of a
holographic c-function defined in terms of entanglement entropy.

%%%%%%%%%%%%%%%%%%%%%%%%%%%%%%%%%%%%%%%%%%%%%%%%%%%%%%%%%%%%%%%%%%%%
\section{Holographic entanglement entropy and a c-function} \label{ent}

Before beginning our holographic analysis, we must first identify a
candidate c-function using entanglement entropy for $d\ge3$. Recall
that \cite{casini2} identifies such a c-function for two-dimensional
quantum field theories as
 \be
c_2= 3\,\ell\,\frac{\partial S}{\partial \ell}
 \labell{dim2}
 \ee
where $S$ is the entanglement entropy for an interval of length $\ell$
on an infinite line. As described above, using the result \reef{twod}
for the entanglement entropy of $d=2$ CFT's, one finds $c_2=c$ at any
fixed points of the RG flows, \ie eq.~\reef{dim2} yields the central
charge of the underlying CFT at the fixed points. We would like to
emulate this construction in higher dimensions. However, one should
recall that in general the entanglement entropy for field theories in
higher dimensions will contain many (non-universal) power law
divergences depending on the geometry of the entangling surface, \eg
see eq.~\reef{entex}. Hence we expect a simple derivative with respect
to some scale characteristic of the entangling surface will typically
yield a result which depends on the cut-off. While there may be various
strategies to avoid this outcome -- see further discussion in section
\ref{discuss} -- here we take the following simple approach: First we
note that, at the fixed points, the power law divergences are geometric
in origin and all but the leading area-law terms vanish if the
geometries of the background and the entangling surface are both flat.
Hence we consider a `strip' or `slab' geometry, where the entangling
surface consists of two parallel flat ($d$--2)-dimensional planes
separated by a distance $\ell$ in a flat background spacetime. The
entanglement entropy (of a CFT) then takes the simple form
\cite{Ryu1,Ryu2}
 \be
S_\mt{CFT}= \alpha_d \frac{H^{d-2}}{\delta^{d-2}} -
\frac{1}{(d-2)\beta_d}\,\C_d\,\frac{H^{d-2}}{\ell^{d-2}}\,,
 \labell{strip}
 \ee
where $\alpha_d$ and $\beta_d$ are dimensionless numerical factors and
$H$ is a(n infrared) regulator distance along the entangling surface --
we assume that $H\gg\ell$. That is, $H^{d-1}$ is the area for each of
the planes comprising the entangling surface and so the first
contribution in eq.~\reef{strip} is simply the usual area law term. The
coefficient of the second finite term is proportional to a central
charge in the underlying $d$-dimensional CFT, which we denote $\C_d$.
Hence we can isolate this central charge by writing
 \be
\C_d=\beta_d\,\frac{\ell^{d-1}}{H^{d-2}}\,\frac{\partial
S_\mt{CFT}}{\partial\ell} \,.
 \labell{cntral}
 \ee
Hence we are naturally lead to consider the quantity
 \be
c_d=\beta_d\,\frac{\ell^{d-1}}{H^{d-2}}\,\frac{\partial
S}{\partial\ell}
 \labell{candid}
 \ee
as a candidate for a c-function along the RG flows, so that
$\tc_d=\C_d$ at the fixed points of the flow. We will identify the
precise value of the coefficient $\beta_d$ with our holographic
calculations below --- see eq.~\reef{cntral2}. Comparing
eqs.~\reef{dim2} and \reef{candid}, we can view the latter expression
as the simplest generalization of the two-dimensional c-function
\reef{dim2} to higher dimensions. At the outset, we wish to say that we
will find below that will only be able to prove that this candidate
c-function actually decreases monotonically along RG flows for
holographic models with Einstein gravity in the bulk. However, another
goal in the following analysis is to connect the behaviour of this
c-function defined using holographic entanglement entropy with the
standard discussions of holographic c-theorems
\cite{gubser,flow2,cthem}. We should also mention that
eq.~\reef{candid} was previously suggested as a c-function in
\cite{Ryu2}.

\subsection{Holographic entanglement entropy on an interval} \label{holoEE}

In this section, we derive some of useful results to evaluate
eq.~\reef{candid} for holographic RG flows in following sections. The
holographic models in sections \ref{one} and \ref{examp1} will be
described by Einstein gravity in the bulk, while we will consider GB
gravity \cite{lovel} in section \ref{three}.

The seminal work of Ryu and Takayanagi \cite{Ryu1,Ryu2} provided a
holographic construction to calculate entanglement entropy. In the
$d$-dimensional boundary field theory, the entanglement entropy between
a spatial region $V$ and its complement $\bar V$ is given by the
following expression in the ($d$+1)-dimensional bulk spacetime:
 \be
S(V) = \frac{2\pi}{\lp^{d-1}}\ \mathrel{\mathop {\rm
min}_{\scriptscriptstyle{v\sim V}} {}\!\!} \left[A(v)\right]
 \labell{define}
 \ee
where $v\sim V$ indicates that $v$ is a bulk surface that is homologous
to the boundary region $V$ \cite{head,furry}. In particular, the
boundary of $v$ matches the `entangling surface' $\partial V$ in the
boundary geometry. The symbol `min' indicates that one should extremize
the area functional over all such surfaces $v$ and evaluate it for the
surface yielding the minimum area.\footnote{We are using `area' to
denote the ($d$--1)-dimensional volume of $v$. If eq.~\reef{define} is
calculated in a Minkowski signature background, any extremal surfaces
are saddle points of the area functional and one should choose the
extremum with the minimum area. However, if one first Wick rotates to
Euclidean signature, the extremization procedure actually corresponds
to finding the global minimum of the area functional.}
Eq.~\reef{define} assumes that the bulk physics is described by
(classical) Einstein gravity and we have adopted the convention:
$\lp^{d-1}=8\pi G_\mt{N}$ Hence the functional which is extremized on
the right-hand side of eq.~\reef{define} matches the standard
expression for the horizon entropy of a black hole. While this proposal
passes a variety of consistency tests, \eg see \cite{Ryu2,head,ent1},
there is no general derivation of this holographic formula
\reef{define}. However, a derivation was recently provided for the
special case of a spherical entangling surface in \cite{casini9}.

In \cite{ent2,ent1}, the above expression \reef{define} was extended to
holographic theories dual to GB gravity \reef{GBAction} in the bulk.
The new prescription still extremizes over bulk surfaces $v$ which
connect to the entangling surface at the asymptotic boundary, however,
the entropy functional to be extremized becomes\footnote{This
expression was motivated by the construction of black hole entropy for
Lovelock gravity appearing in \cite{jacobson}. We note that when
evaluated on a general surface this functional will not match the Wald
entropy \cite{WaldEnt}. However, the two agree when evaluated on a
stationary black hole horizon.}
 \bea
S &\,=\,&  {2\pi \ov \lp^{d-1}} \int_v d^{d-1}x \, \sqrt{h} \, \left[ 1
+ {2 \,\lambda L^2\ov (d-2)(d-3)} \mathcal{R} \ri] \nonumber \\
\ & \ & \qquad \qquad \qquad \qquad + {4 \pi \ov \lp^{d-1}}
\int_{\partial v} d^{d-2}x\sqrt{\tilde{h}} {2 \, \lambda\, L^2 \ov
(d-2)(d-3)}
 \, \mathcal{K}\,.
 \labell{gbstrip}
 \eea
Here, $h$ ($\tilde h$) is the induced metric on (the boundary of) the
bulk surface $v$, $\mathcal{R}$ is the Ricci scalar of this induced
geometry and $\mathcal{K}$ is the extrinsic curvature of the boundary
$\partial v$ at the asymptotic cut-off surface. Note that we only apply
this expression for $d\ge4$ since it is only for these dimensions that
the GB interaction \reef{GBterm} contributes to the gravitational
equations of motion. Of course, if we set $\lambda=0$ in the above
expression, it reduces to $2\pi A(v)/ \lp^{d-1}$ and we recover
eq.~\reef{define}. Note that the extrinsic curvature term in
eq.~\reef{gbstrip} plays the role of a `Gibbons-Hawking' surface term
to ensure that the variational principle is consistent.

\FIGURE{
\includegraphics[width=10cm,height=6cm]{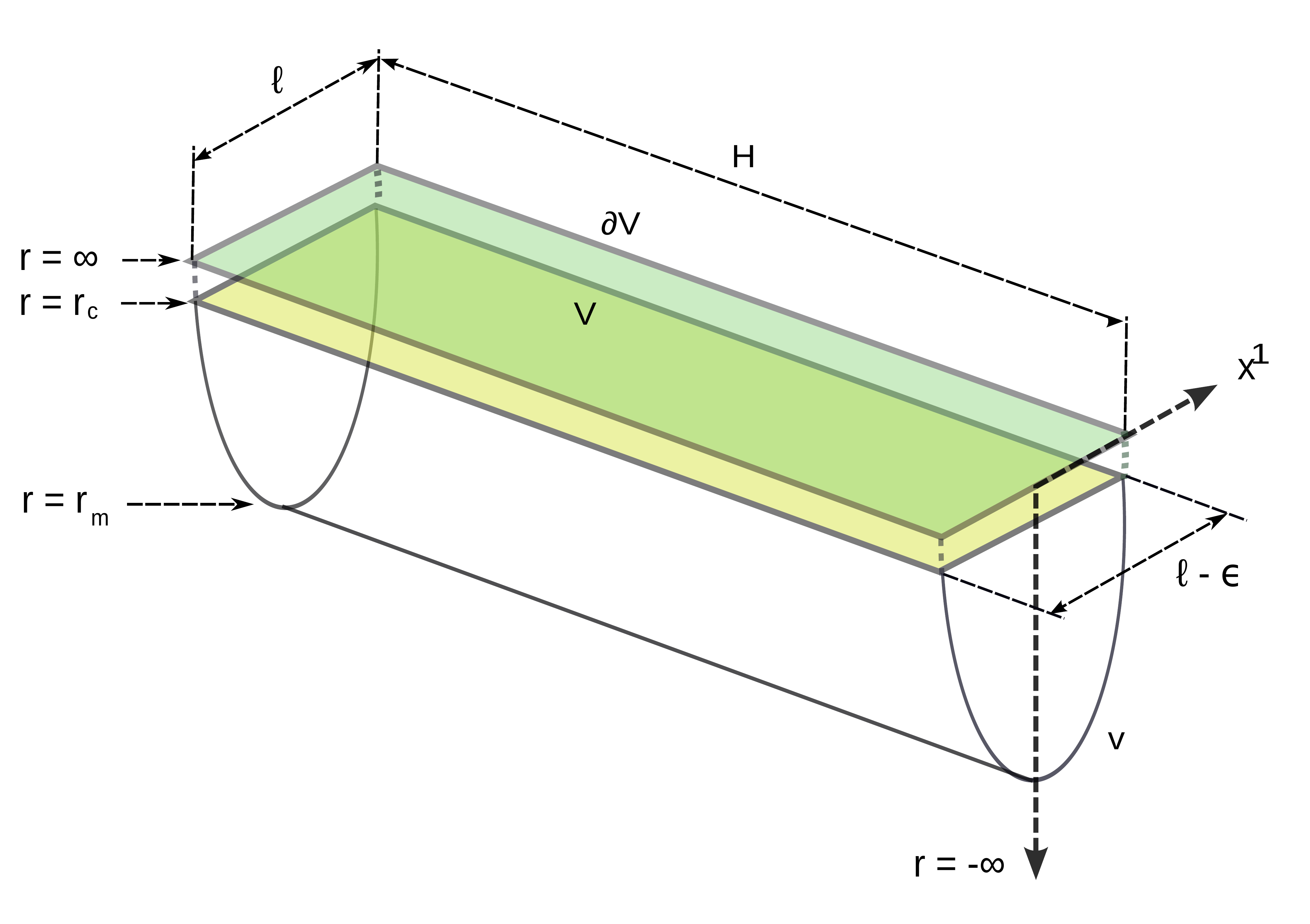}
\caption{(Colour Online) The strip $V$ in the asymptotic boundary, with
the minimal surface $v$ in the bulk ending on $\partial V$. The
entangling surface $\partial V$ consists of two flat (hyper)planes
positioned at $x^1=\ell/2$ and $x^1=-\ell/2$. A regulator length $H$ is
introduced to limit the extent of these planes along the remaining
directions.} \label{house}}
Now let us begin to consider evaluating the holographic entanglement
entropy for a general RG flow. As in the previous section, we assume
the bulk metric takes the form given in eq.~\reef{metric0}. Then the
boundary geometry is simply flat space and we define the entangling
surfaces as follows: First recall that the entangling surface divides a
Cauchy surface (\eg the constant time slice, $x^0=t=0$) into two
regions. As described above, we wish to consider an interval of length
$\ell$ and so we introduce two flat (hyper)planes at $x^1=\ell/2$ and
$x^1=-\ell/2$, as shown in figure \ref{house}. We also introduce a
regulator length $H$ to limit the size of the two planes along the
$x^{2,3,\cdots,d-1}$ directions, \eg we can imagine the boundary is
periodic in these directions with length $H\gg\ell$. Hence the area of
either plane is $H^{d-1}$, as described at eq.~\reef{strip}. In
calculating the holographic entanglement entropy, we consider bulk
surfaces that end on the entangling surface as $r\to \infty$, as shown
in figure \ref{house}. With the `slab' geometry described here, the
radial profile of these surfaces will only be a function of the
coordinate $x=x^1$. We will write the profile as $r=r(x,w=\l)$ where
$w$ indicates the width of the interval which sets the boundary
condition, \ie in the present case, $x \to \l/2$ as $r \to \infty$. Of
course, the holographic calculations are only well-defined if we
introduce an asymptotic cut-off surface as some $r=r_c$. The position
of this surface is related to a short distance cut-off in the boundary
theory, \ie $r_c=\tL\,\log(\tL/\delta)$. The radial profile will define
another useful UV scale $\epsilon$ with
$r_c=r(x=(\ell-\epsilon)/2,w=\l)$, \ie the profile intersects the
cut-off surface $r=r_c$ at $x=(\ell-\epsilon)/2$. Another useful scale
in the bulk surface is the minimal radius $r_m$ which it reaches in the
bulk, which appears as $r_m=r(x=0,w=\l)$.

Given the background metric \reef{metric0} and our ansatz $r(x,\l)$ for
the profile of the bulk surface, we find that eq.~\reef{gbstrip}
reduces to the following simple expression
 \be
S ={2\pi \ov \lp^{d-1}} \int_m d^{d-1}x \, \frac{e^{(d-2)A} \left(e^{2
A}+\left(1+2 \,\lambda L^2 \, A'{}^2\right) \dr^2 \right)}{\sqrt{e^{2
A}+\dr^2}}
%\equiv {2\pi \ov \lp^{d-1}} \int_m d^{d-1}x \,\mathcal{L}
\,. \labell{functional}
 \ee
where $\dr=\partial r/\partial x$ and $A'=dA/dr$. Now one may treat the
above expression as an action which is varied to find a second-order
differential equation to determine the profile $r(x,\l)$. However,
since the integrand above has no explicit dependence on the coordinate
$x$, the following is a conserved quantity along the radial
profile\footnote{If we denote the integrand in eq.~\reef{functional} as
$\mathcal{L}$, then $K_d(\l)^{-1}=\mathcal{L} - {d \mathcal{L} \ov d
\dr} \,\dr$.}
 \be
{K_d(\l)} \,\equiv\,  {e^{-d A} \,(e^{2A}+\dr^2)^{3/2}\ov  e^{2A} +(1 -
2\, \lambda L^2 \, A'{}^2)\, \dr^2  }\,. \labell{conserved}
 \ee
This leaves us with a first-order equation for the profile, which
should be easier to solve. In principle then, our goal is to solve for
$r(x,\l)$ in a given holographic RG flow geometry, \ie for a specific
conformal factor $A(r)$, and then substitute the solution back into
eq.~\eqref{functional} to calculate the entanglement entropy.

Before going on to consider the entanglement entropy and c-function for
RG flow geometries, let us first examine the results when the bulk
geometry is simply AdS space, \ie at a fixed point of the flow where
the boundary theory is conformal. Recall that for the AdS vacuum
$A(r)=r/\tL$. Let us begin by setting $\lambda=0$ and considering the
results for Einstein gravity in the bulk.\footnote{Note that in this
case, the AdS curvature is given by simply $\tL=L$.} The case of
three-dimensional AdS or a $d=2$ boundary CFT is special since the
entanglement entropy yields a logarithmic UV divergence
 \be
S_\mt{CFT} \;=\; {4 \, \pi \tL \ov \lp}\, \log \left({\l \ov \delta}
\ri)\,. \labell{d2ads}
 \ee
If we recall that the central charge of the boundary CFT is given by
$c=12\pi\,\tL/\lp$, we see that this expression precisely reproduces
the expected result \reef{twod} for the entanglement entropy of a
two-dimensional CFT. Next turning to Einstein gravity with $d\ge3$, the
entanglement entropy for the interval is given by \cite{Ryu1}
 \be
S_\mt{CFT}\; =\;\frac{4\pi}{d-2}\, { \tL^{d-1} \ov  \lp^{d-1}} \left(
{H \ov \delta}\right)^{d-2} - {2^{d} \, \pi^{(d+1)/2} \,  \ov (d-2)
 }\left( { \Gamma({d \ov 2(d-1)}) \ov \Gamma({1 \ov 2(d-1)})}
\right)^{d-1}\,{ \tL^{d-1} \ov  \lp^{d-1}}\, \left( H \ov \l
\right)^{d-2}\,.
 \labell{eqn12}
 \ee
Here we see the general structure given in eq.~\reef{strip} with two
terms, a power law divergence proportional to $(H/\delta)^{d-2}$ and a
finite contribution proportional to $(H/\l)^{d-2}$. Next for
$\lambda=0$, both of the central charges in eqs.~\reef{effectc} and
\reef{effecta} are identical and we use this fact to define $\C_d$ in
eq.~\reef{strip} for Einstein gravity: $ \C_d(\lambda=0)\equiv\pi^{d/2}
/\Gamma(d/2)\, (\tL/\lp)^{d-1}=\ct=\ads$. As described previously then,
we can extract this central charge from the above entanglement entropy
using eq.~\reef{cntral}, which yields
 \be
\C_d =\beta_d\,\frac{\ell^{d-1}}{H^{d-2}}\,\frac{\partial
S_\mt{CFT}}{\partial\ell}\quad {\rm with}\ \ \beta_d=
\frac1{\sqrt{\pi}\, 2^{d}\, \Gamma(d/2)  } \left(\Gamma({1 \ov
2(d-1)})\ov { \Gamma({d \ov 2(d-1)})} \right)^{d-1}\,.
 \labell{cntral2}
 \ee
Hence we have identified the precise value of $\beta_d$ (for $d\ge3$)
which appears as the coefficient in eq.~\reef{candid} of the
c-function.

Finally let us apply the above formulae to calculate holographic
entanglement entropy with the strip geometry for the boundary CFT dual
to the AdS vacuum in GB gravity \reef{GBAction}. To simplify the final
results, it is convenient to first treat $r$ as the independent
variable, in which case to fix the profile of the bulk surface, we must
determine $x(r)$. Next we choose a new radial coordinate $\tau =
K^{-1/(d-1)}\,e^{-r/\tL}$ and define $h\equiv e^{2r/\tL}\,(\partial_r
x)^2$. Note that as $r\to\infty$, $\tau\to0$ and further one can show
at $r=r_m$, $\tau=1$. Now with these choices, eq.~\eqref{conserved}
becomes
 \be
\frac{\tau^{d-1}\,(1+h)^{3/2}}{\sqrt{h}\,(1-2 \lambda f_\infty +
h)}=1\,.
 \label{new38}
 \ee
In general, this equation yields three roots for $h(\tau)$ and the
relevant solution is the real root which can be continuously connected
to the $\lambda=0$ solution: $h=\tau^{2(d-2)}/(1-\tau^{2(d-2)})$. Now
it is straightforward to see that the entanglement entropy
\reef{functional} can be written as
 \bea
S_\mt{CFT} &\; =\;& {4\pi\ov d-2}\left(1+2\lambda f_\infty \right) {
\tL^{d-1} \ov  \lp^{d-1}} \left( {H \ov \delta}\right)^{d-2}
-\frac{4\pi }{d-2}\left(1+2\lambda f_\infty \right){ \tL^{d-1} \ov
\lp^{d-1}}
 \left( {H \ov \l} \right)^{d-2} \nonumber \\
&&\qquad \times\ I^{d-2}\,\Bigg[ 1+(d-2) \int_0^1
\frac{d\tau}{\tau^{d-1}} \Bigg(1- {1+2\lambda f_\infty + h \ov
\left(1+2\lambda f_\infty \right) \sqrt{1+h}} \Bigg)\Bigg] \,,
 \labell{gbee}
 \eea
where $I\equiv \int_0^1 d\tau \sqrt{h}$. Then applying
eq.~\reef{cntral2}, we can express the central charge in the finite
contribution as\footnote{Note that analogous results were given for the
case $d=4$ in \cite{ent2}. However, we note that the calculations
presented there did not include the `Gibbons-Hawking' surface term in
eq.~\reef{gbstrip} and hence their expressions do not match those
presented here. However, we have verified numerically that the
effective central charge in \cite{ent2} agrees with eq.~\reef{gbcharge}
when $d=4$. We also observe that the leading divergent term in
eq.~\reef{gbee} is proportional to $\ct$ while without the
`Gibbons-Hawking' term, this term is proportional to $\ads$.}
 \bea
\C_d(\lambda) &\;=\;& 4\pi\beta_d \left(1+2\lambda f_\infty \right){
\tL^{d-1} \ov \lp^{d-1}} \,I^{d-2}\,\Bigg[ 1+(d-2) \int_0^1
\frac{d\tau}{\tau^{d-1}} \Bigg(1- {1+2\lambda f_\infty + h \ov
\left(1+2\lambda f_\infty \right) \sqrt{1+h}} \Bigg)\Bigg]
 \,.
 \labell{gbcharge}
 \eea
Regrettably, we do not have a closed analytic expression for
$\C_d(\lambda)$ in terms of the two central charges $\ct$ and $\ads$.
Hence we have numerically evaluated the above expression and plotted
$\C_d(\lambda)/\ads$ as a function of $\ct/\ads$ in figure \ref{cplot}a
for several values of $d$. Note that in this figure,
$\C_d(\lambda)/\ads=1$ at $\ct/\ads=1$ for all of the values of $d$
since this corresponds to $\lambda=0$ or Einstein gravity in the bulk.
From these curves, we can infer that $\C_d(\lambda)$ is a complicated
nonlinear function of both $\ct$ and $\ads$.  We can also illustrate
this fact as follows: In the vicinity of $\lambda\simeq0$ or
$\ct\simeq\ads$, we can make a linearized analysis of
eq.~\reef{gbcharge} to find
 \be
\C_d(\lambda) = \C_d(0)\left(1+ {3(d-1)\ov 2}\lambda + O(\lambda^2)\,
\right)= \C_L(\lambda) + O(\lambda^2)
 \labell{adscfun}
 \ee
where
 \be
\C_L(\lambda)\equiv\frac{(d-1)(d-2)}2\, \ct + \frac{d (d-3)}2\, \ads
\,.
 \labell{cline}
 \ee
Here, we have defined $\C_L(\lambda)$ as the linear combination of the
two central charges in eqs.~\reef{effectc} and \reef{effecta} which
yields an $O(\lambda)$ expansion which precisely matches that for
$\C_d(\lambda)$.
%Note that the result in eq.~\reef{adscfun} shows that for curves plotted
%in figure \ref{cplot}a, the slope at $\ct/\ads=1$ increases as $d$ grows.
Next, we consider the ratio of $\C_d(\lambda)$ and $\C_L(\lambda)$ over
the full (physical) range of $\lambda$. Since $\C_L(\lambda)$ can
vanish in this range, it is convenient plot the ratio
$\C_L(\lambda)/\C_d(\lambda)$ as a function of $\ct/\ads$, as shown in
figure \ref{cplot}b. This figure illustrates even more dramatically our
previous observation that $\C_d(\lambda)$ is a complicated nonlinear
function of both $\ct$ and $\ads$. At this point, let us add that since
$\C_d(\lambda)\ne\ads$, the central charge identified in \cite{cthem}
as satisfying a c-theorem, we might not expect that our new effective
central charge $\C_d(\lambda)$ will always flow monotonically in
holographic RG flows for general $\lambda$.
\FIGURE[!ht]{
\begin{tabular}{cc}
\includegraphics[width=0.5\textwidth]{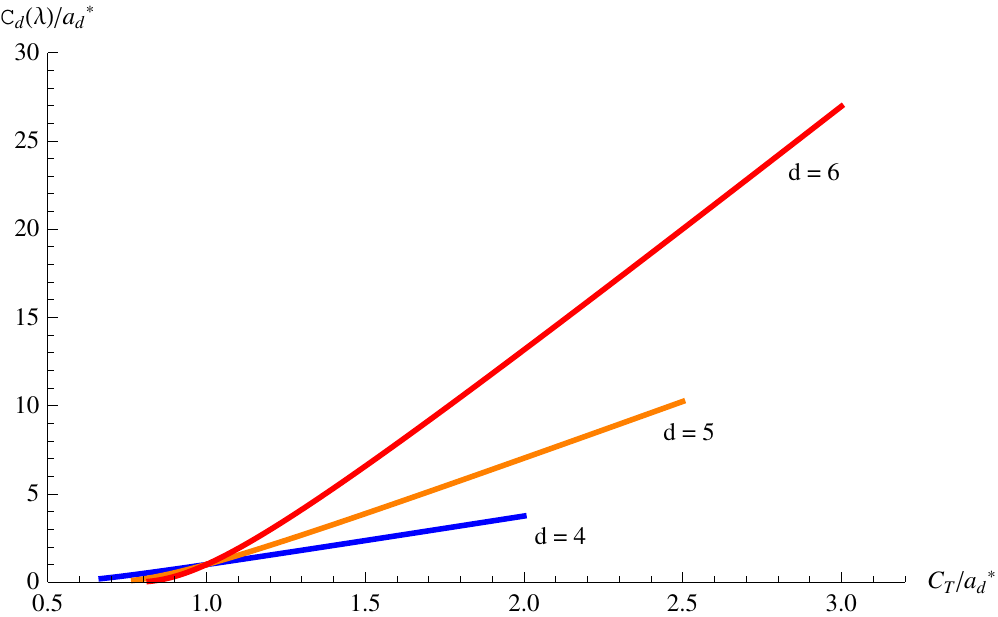}&
\includegraphics[width=0.5\textwidth]{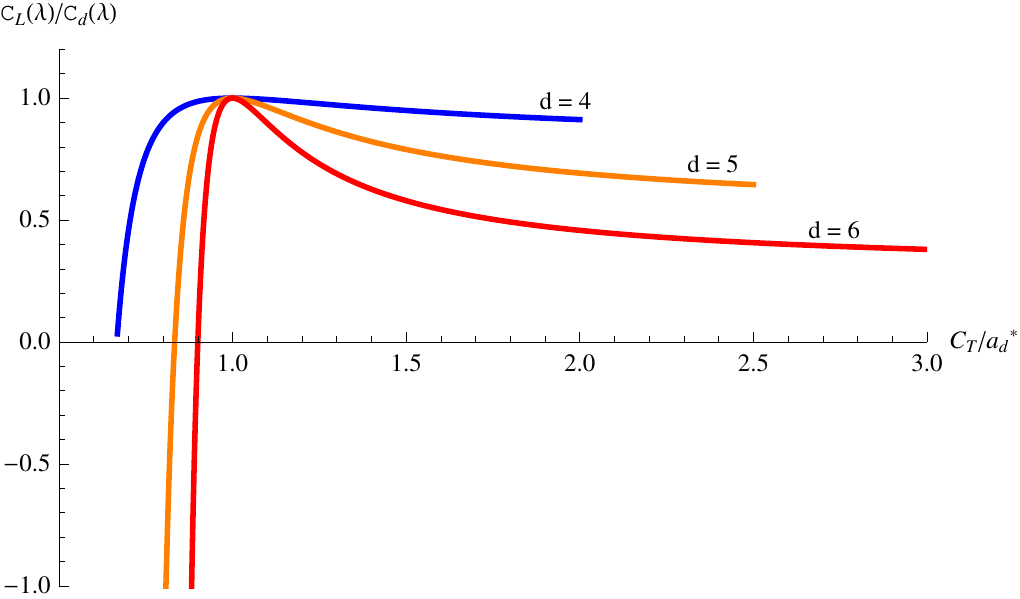}\\
(a) & (b)
\end{tabular}
\caption{(Colour Online) Panel (a) plots $\C_d(\lambda)/\ads$ as a
function of $C_T/\ads$ for GB gravity, while panel (b) is a plot of
$\C_L(\lambda)/\C_d(\lambda)$ as a function of $C_T/\ads$. Note that
both $\C_d(\lambda)/\ads=1=\C_L(\lambda)/\C_d(\lambda)$ at $C_T/\ads=1$
which corresponds to Einstein gravity in the bulk (\ie $\lambda=0$).
Each curve runs over the physically allowed range of $C_T/\ads$ for the
given value of $d$ --- see eq.~\reef{limitsca} for more details.}
\label{cplot}}
%%%%%%%%%%%%%%%%%%%%%%%%%%%%%%%%%%%%%%%%%%%%%%%%

\section{Holographic flow of c-function with Einstein gravity} \label{one}

In this section, we examine the behavior of the c-function
\eqref{candid} in a general holographic RG flow dual to Einstein
gravity. We will first discuss the flow of the c-function in $d=2$ and
then generalize it to arbitrary dimensions.

For $d=2$, the bulk theory is three-dimensional Einstein gravity
coupled to, \eg a scalar field with a nontrivial potential, as
described in section \ref{review}. Our holographic expression
\reef{functional} for the entanglement entropy of a strip can be
written as
 \be
S \,=\, {4\pi \ov \lp} \int_0^{\frac{\l-\epsilon}2} dx \,
\sqrt{\dr^2+e^{2A}}\,. \labell{onex1}
 \ee
Note that the integration above runs over half of the range, \ie $x \in
[\,0,(\l-\epsilon)/2\,]$. Further the conserved charge
\eqref{conserved} simplifies to
 \be
K_2(\l)\,=\,e^{-2A(r)} \, \sqrt{\dr^2+e^{2A}}\,. \labell{onex2}
 \ee
To calculate $dS/d\l$, we note that $\dr=\dr(x,\l)$ and $A=A(r(x,\l))$,
\ie the profile of the extremal surface implicitly depends on the strip
width $\l$. If we vary $S$ with respect to $\l$, keeping the UV cut-off
$r=r_c$ fixed, we will get two contributions: one coming from change in
the limits of the integration and second from change in the solution
$r(x,\l)$. We write them as
 \be
{dS \ov d\l} \,=\, {4 \, \pi \ov \lp} {1 \ov \sqrt{\dr^2 + e^{2A(r)}}}
\left[{1\ov 2} \left(1- {d\epsilon \ov d\l} \ri) (\dr^2+e^{2A(r)}) +
\dr \, {\partial r \ov \partial \l} \ri] \Bigg|_{x={\l-\epsilon \ov 2}}
\,,
 \labell{onex3}
 \ee
where we have used the equation of motion for $r(x,\l)$
 \be
\ddot{r} - A'\, \left(2 \dr^2 +  e^{2 A} \right)\,=\,0\,,
 \ee
to cancel the bulk contribution. Since the UV cut-off $r_c$ is fixed
while performing the variation, we get some extra constraints between
$\dr=\partial_x r$ and $r'=\partial_\l r$ at the asymptotic boundary.
Taking variation of relation $r({\l-\epsilon \ov 2},\l)=r_c$ with
respect to $\l$, we get
 \be
\left[{\dr(x,\l) \ov 2} \left( 1 - {d \epsilon \ov d\l}\ri) + {\partial
r(x,\l) \ov \partial \l} \right]_{x={\l-\epsilon \ov 2}} \,=\,0\,.
 \labell{onex4}
 \ee
Substituting this relation, as well as eq.~\eqref{onex2}, into
eq.~\eqref{onex3} gives us following expression for $dS/ d \l$:
 \be
{dS \ov d \l} \,=\, -{4 \pi \ov \lp \, K_2(\l)}{1 \ov \dr } {\partial r
\ov \partial \l} \bigg|_{x={\l-\epsilon \ov 2}}\,.
 \labell{onex5}
 \ee
In the above relation, the partial derivatives of $r(x,\l)$ are
evaluated near the asymptotic boundary. To further simplify this
expression, we use the  Fefferman-Graham expansion near the boundary
\cite{fefferman}. In terms of the radial coordinate $r$, this expansion
takes the form \cite{calc}
 \be
ds^2 \,=\, dr^2 + e^{{2r/{\tilde{L}}}} \, f(r)\, \eta_{ij}\, dx^i\,dx^j
\,,
 \labell{feffer}
 \ee
where
 \be
f(r) \,=\, 1 + a_1 \phi_0^2\,e^{-2\alpha r/\tilde{L}} + a_3
\phi_0^3\,e^{-3\alpha r/\tilde{L}} + \cdots \;.
 \labell{feffer1}
 \ee
In this expansion, $\tilde{L}$ is the AdS radius in the UV region (\ie
as $r\to\infty$) and $\alpha=d-\Delta$ where $\Delta$ is the conformal
weight of the operator dual to the bulk scalar field. Near the
boundary, the coordinate $r$ is very large and hence it is sufficient
to work with only the leading order term in the expansion
\eqref{feffer1}. Although $A(r)$ has a complicated profile deep inside
the bulk, near the boundary it will have the simple form
 \be
A(r)\,=\,{r / {\tilde{L}}}\,. \labell{onex6}
 \ee
For this $A(r)$, eq.~\reef{onex2} can be re-expressed as the following
equation of motion: $dx/dr=e^{-2r/\tilde{L}} /\sqrt{K_2^2 -e^{-2r
/\tilde{L}}}$. The latter is easily integrated to yield the following
solution
 \be
x-{\l \ov 2}\,=\, \tilde{L}\sqrt{K_2^2-e^{-2r/\tilde{L}}} -\tilde{L}K_2
\,. \labell{onex7}
 \ee
where the integration constant was chosen so that $x\to\l/2$ as
$r\to\infty$. Next we differentiate the above solution with respect to
$x$ and $\l$ to find $\dr$ and $\partial r/
\partial \l$, treating that $K_2(\l)$ as a function of
$\l$ -- see appendix \ref{two} for further details. Taking the limit
$r\to \infty$ in ratio of $\partial r/ \partial \l$ and $\dr$ appearing
in eq.~\reef{onex5}, we find that
 \be
{1 \ov \dr} {\partial r \ov \partial \l} \bigg|_{x=\l/2} \,=\, -{1 \ov
2}\,.
 \labell{onex8}
 \ee
This relation not only simplifies eq.~\eqref{onex5} but also ensures
that the first derivative of $S$ is indeed finite for all RG flow
solutions. Using this relation in eq.~\eqref{onex5}, we arrive at
following elegant form of the c-function \eqref{dim2} for arbitrary RG
flow backgrounds:
 \be
{c}_2\,=\,3\,\l\, {dS \ov d \l} \,=\, {6 \pi \ov \lp}{\l \ov K_2(\l)
}\,. \labell{onex9}
 \ee

The next step is to show that this c-function increases monotonically
along holographic RG flows. Implicitly, the extremal bulk surfaces $v$
on which we are evaluating the Ryu-Takayanagi formula \reef{define}
extend to infinite $r$ at $x=\pm\l/2$ and pass through a minimum
$r=r_m$ at $x=0$. The latter radius gives us an indication of which
degrees of freedom the entanglement entropy is probing, \ie for smaller
values of $r_m$, we expect the entropy and $c_2$ responds more to the
IR structure of the RG flow. Hence in the following, we will study
behavior of ${c}_2$ as a function of the turning point radius $r_m$ and
we wish to establish the `c-theorem' as $dc_2/dr_m\ge0$ -- at least for
background geometries that satisfy appropriate constraints.

Comparing to the field theory construction of \cite{casini2}, we note
that there the c-theorem was formulated as $dc_2/d\l\le0$. Naively,
this result matches with the holographic inequality which we wish to
establish since we expect that as the width of the strip increases, the
minimal area surface will explore deeper regions in the bulk geometry.
The two inequalities would be rigorously connected if we could prove a
second inequality $d\l/dr_m\le0$ for consistent holographic models.
However, as we will see in the next section, in fact this inequality
does not hold for all extremal surfaces. However, we will still find
$dc_2/d\l\le0$ in all cases of interest. The violations of the previous
inequality are associated with unstable saddle-points which do not
contribute to the physical entanglement entropy. Hence, in section
\ref{examp1}, we will find that the behaviour of the entanglement
entropy in general holographic RG flows provides a richer story than
might have been naively anticipated.

Returning to the flow of the c-function, we note that at the minimum of
the bulk surface, we will have $r(0,\l)=r_m$ and $\dr(0,\l)=0$. Hence
considering eq.~\eqref{onex2} at this turning point, we find
 \be
K_2(r_m)\,=\,e^{-A(r_m)}\,. \labell{onex10}
 \ee
Here it is natural to treat this constant of the motion as a function
of $r_m$, rather than $\l$. We will also work with width of the strip
$\l$ as function of $r_m$. Then combining eqs.~\eqref{onex9} and
\eqref{onex10} yields
 \be
{d{c}_2 \ov d r_m } \,=\, {6 \pi \ov \lp K_2(r_m)}\left( {d\l \ov d
r_m} + A'(r_m) \, \l  \ri)\,.
 \labell{onex11}
 \ee
Now to express $\l$ in terms of $r_m$, we begin with the relation
 \be
{\l\ov 2} \,=\, \int_0^{\l/2} dx \,=\, \int_{r_m}^\infty {dr \ov \dr}
\,=\, \int_{r_m}^\infty dr {e^{-2A} \, A'  \ov \sqrt{K_2^2 - e^{-2A}}}
{1\ov A'}\,.
 \labell{onex12}
 \ee
Here in the final expression we have used eq.~\eqref{onex2}. Now above,
we will apply integration by parts using
 \be
\int dr\,\frac{e^{-2A}A'}{\sqrt{K_2^2-e^{-2A}}}=\sqrt{K_2^2-e^{-2A}}
\,.
 \labell{byparts}
 \ee
to find that
 \be
{\l} \,=\, 2\tL K_2 + 2\int_{r_m}^{\infty} dr {A'' \ov A'{}^2}
\sqrt{K_2^2 - e^{-2A}}\,, \labell{onex13}
 \ee
where we have used eq.~\eqref{onex6} to evaluate $A'(r)$ at $r=\infty$.
Further we can differentiate this expression with respect to $r_m$ to
get
 \be
{d \l \ov d r_m} \,=\, - 2 \tL A'(r_m) \,K_2 - 2 A'(r_m)
\int_{r_m}^\infty dr {A'' \ov A'{}^2} {K_2^2 \ov \sqrt{K_2^2 -
e^{-2A}}}\,. \labell{onex14}
 \ee
Now substituting eqs.~\eqref{onex13} and \eqref{onex14} into
eq.~\eqref{onex11}, we find
 \bea
{d {c}_2 \ov d r_m } & \,=\, & - {12 \, \pi A'(r_m)  \ov \lp\,
K_2(r_m)} \int_{r_m}^\infty dr {A'' \ov A'{}^2} {e^{-2A} \ov
\sqrt{K_2^2 - e^{-2A}}}\,
 \nonumber \\
\ & \,=\, & - {12 \, \pi A'(r_m)  \ov \lp\, K_2(r_m)} \int_{0}^\l dx\,
{A'' \ov A'{}^2} \,,
 \labell{onex15}\\
\ & \,=\, & - {12 \, \pi A'(r_m)  \ov \lp\, K_2(r_m)} \int_{0}^\l dx\,
{1 \ov A'{}^2}\,\left(T^t{}_t-T^r{}_r\right)\ge0 \,.
 \nonumber
 \eea
In the second line, we have used eq.~\eqref{onex2} to convert the
integration over $r$ to one over $x$. In the last line, we have used
Einstein's equations to replace $A''$ by the components of the stress
tensor. As for the discussion of holographic c-theorems in section
\ref{review}, the final inequality assumes that the bulk matter fields
driving the holographic RG flow satisfy the null energy condition. The
latter ensures that the integrand is negative. The overall inequality
also requires $K_2(r_m)>0$ and $A'(r_m)>0$. The first condition is
obvious from eq.~\reef{onex10} while the second can be established as
follows: Given the null energy condition, it follows that $A'' \leq 0$
which means that $A'$ is everywhere a decreasing function of radial
coordinate $r$. Implicitly, we are assuming the bulk geometry
approaches AdS space asymptotically, \ie the dual field theory
approaches a conformal fixed point in the UV. Hence with $r\to\infty$,
we see the minimal value of $A'$ is $A'=1/\tilde{L}$, where $\tilde{L}$
is the asymptotic AdS scale. Since this minimal value is positive, it
must be that $A'$ is positive everywhere along the holographic RG flow.
Hence $d{c}_2/d r_m$ is positive and our two-dimensional c-function
increases monotonically along the RG flow if the bulk matter satisfies
the null energy condition.

We now turn to proving the monotonic flow of the c-function
\eqref{candid} for higher dimensions. The required analysis is a
straightforward extension of the above calculations with $d=2$. In
particular, one finds that eq.~\eqref{onex9} generalizes to
 \be
{c}_d \,=\, {2 \, \pi \,\beta_d \ov \lp^{d-1}} { \l^{d-1} \ov K_d(\l)}
\,,
 \labell{rex1}
 \ee
with $d$ boundary dimensions. The conserved quantity \eqref{conserved}
is now given by
 \be
K_d \,=\, e^{- d A(r)}  \sqrt{\dr^2 + e^{2A(r)}} \,. \labell{twox3a}
 \ee
We have relegated the detailed derivation of eq.~\reef{rex1} to
appendix \ref{two}. However, we can see from this result that all the
complexities of determining the c-function boil down to evaluating the
conserved charge \reef{twox3a} for the minimal area surface. We might
note that we can evaluate this expression at the minimal radius (where
$\dr=0$) to find
 \be
K_d(r_m)=e^{-(d-1)A(r_m)}\,,
 \labell{onex10a}
 \ee
which generalizes eq.~\reef{onex10} to general $d$.

Combining eqs.~\eqref{rex1} and \eqref{onex10a}, we further find
 \be
{d{c}_d \ov dr_m} \,=\, {2 \,(d-1) \pi \,\beta_d \, \l^{d-2} \ov K_d}
\, \left( \frac{d\l}{dr_m} + A'(r_m) \,\l \ri)\,. \labell{rex3}
 \ee
To express $\l$ in terms of $r_m$, eq.~\reef{onex12} now becomes
 \be
{\l \ov 2} \,=\, \int_{0}^{\l/2} dx\,=\,\int_{r_m}^{\infty} {dr \ov
\dr} \,=\, \int_{r_m}^{\infty} dr {A' \, e^{-d A(r)} \ov \sqrt{K_d^2 -
e^{-2(d-1)A}}} {1 \ov A'}\,, \labell{rex4}
 \ee
where the last expression follows by using eq.~\eqref{twox3a}. To make
further progress, we observe that
 \bea
&&\int dr\,{A' \, e^{-d A(r)} \ov \sqrt{K_d^2 - e^{-2(d-1)A}}}= -{e^{-d
A} \ov d \,K_d} \  {}_2F_{1}\!\left[{1 \ov2 }, {d \ov 2(d-1)}; {3d-2
\ov 2(d-1)}; {e^{-2(d-1)A(r)} \ov K_d^2}\ri]
 \labell{byparts2}\\
&&\quad=e^{-(d-2)A}\sqrt{K_d^2-e^{-2(d-1)A}}-\frac{e^{-(d-3)A}\,K^2_d}{\sqrt{
K_d^2-e^{-2(d-1)A}}} \  {}_2F_{1}\!\left[{1 \ov2 }, -{d-2 \ov 2(d-1)};
{d \ov 2(d-1)}; {e^{-2(d-1)A(r)} \ov K_d^2}\ri] \,.
 \nonumber
 \eea
We have presented the second expression above to illustrate that this
result is simply an extension of eq.~\eqref{byparts} for general $d$
but in the following, we will use the more compact expression given in
the first line. With eq.~\reef{byparts2}, we can integrate by parts in
eq.~\reef{rex4} to find
 \bea
\l
%& \,=\,& -2 \left[ {e^{-d A} \ov d \, A'\,K_d} \ {}_2F_{1}\!\left[{1
%\ov2 },{d \ov 2(d-1)}; {3d-2 \ov 2(d-1)}; {e^{-2(d-1)A(r)} \ov K_d^2} \ri]
% \ri]_{r_m}^{\infty} \nonumber \\
%
%& & \qquad \qquad - 2\int_{r_m}^{\infty}  dr {A'' \ov A'{}^2} {e^{-d A} \ov d \,K_d} \
%{}_2F_{1}\!\left[{1 \ov2 }, {d \ov 2(d-1)}; {3d-2 \ov 2(d-1)}; {e^{-2(d-1)A(r)} \ov K_d^2} \ri]
%\nonumber \\
%
& \,=\, & {2 \sqrt{\pi} \, K_d^{1/(d-1)} \ov d A'(r_m) } \, {\Gamma \left({3d-2 \ov 2(d-1)}\ri)
 \ov \Gamma \left({2d-1 \ov 2(d-1)}\ri) } \labell{rex5} \\
& & \qquad \qquad - 2 \int_{r_m}^{\infty}  dr {A'' \ov A'{}^2} {e^{-d
A} \ov d \,K_d} \ {}_2F_{1}\!\left[{1 \ov2 }, {d \ov 2(d-1)}; {3d-2 \ov
2(d-1)}; {e^{-2(d-1)A(r)} \ov K_d^2} \ri]\,.
 \nonumber
 \eea
Differentiating this result with respect to $r_m$ and making various
simplifications yields
 \bea
{d \l \ov d r_m } &\,=\, & - {2\sqrt{\pi} \, K_d^{1/(d-1)} \ov d } \,
{\Gamma \left({3d-2 \ov 2(d-1)}\ri) \ov \Gamma \left({2d-1 \ov
2(d-1)}\ri)} - 2 {A'(r_m) } \int_{r_m}^{\infty} dr \, {A'' \ov A'^2}
{ e^{-d A} \ov \sqrt{K_d^2 - e^{-2(d-1)A}} } \labell{rex6} \\
&  & \quad + 2 A'(r_m) \int_{r_m}^{\infty}  dr \, {A'' \ov A'^2}\, {
e^{-d A} \ov d \,K_d} \  {}_2F_{1}\!\left[{1 \ov2 }, {d \ov 2(d-1)};
{3d-2 \ov 2(d-1)}; {e^{-2(d-1)A(r)} \ov K_d^2} \ri]\,.
 \nonumber
 \eea
Now substituting eqs.~\eqref{rex5} and \eqref{rex6} into
eq.~\eqref{rex3}, we find
 \bea
{d {c}_d \ov d r_m} &\,=\,&-{ 4\, \pi (d-1) \beta_d \, \l^{d-2}
\,A'(r_m) \ov \lp^{d-1}\,K_d(r_m)} \int_{r_m}^{\infty}  dr \, {A'' \ov
A'^2} {
e^{d A} \ov \sqrt{K_d^2 - e^{-2(d-1)A}}} \nonumber \\
 \ & \,=\, & -{ 4\, \pi (d-1) \beta_d \, \l^{d-2}\,A'(r_m) \ov
 \lp^{d-1}\,K_d(r_m) } \int_{0}^{\l}  dx \, {A'' \ov A'^2}
 \labell{twox19}\\
\ & \,=\, & -{ 4\, \pi  \beta_d \, \l^{d-2}\,A'(r_m) \ov
 \lp^{d-1}\,K_d(r_m) } \int_{0}^{\l}  dx \, {1 \ov A'^2} \,\left(T^t{}_t-
 T^r{}_r\right)\ge0 \,.
 \nonumber
 \eea
The steps here are essentially the same as in our analysis of
eq.~\reef{onex15} with $d=2$. The key requirement for the final
inequality to hold is that the bulk matter fields driving the
holographic RG flow must satisfy the null energy condition. With this
assumption then, $d{c}_d/d r_m$ is positive and our $d$-dimensional
c-function increases monotonically along the RG flow for holographic
boundary theories dual to Einstein gravity in the bulk.

%%%%%%%%%%%%%%%%%%%%%%%%%%%%%%%

\section{Explicit geometries and Phase transitions} \label{examp1}

In this section, we consider some simple bulk geometries describing
holographic RG flows. This allows us to explicitly demonstrate that the
c-function \eqref{candid} indeed flows monotonically for boundary field
theories dual to Einstein gravity. However, we will also find that for
some RG flows, there is a `first order phase transition' in the
entanglement entropy as the width of the strip $\l$ passes through a
critical value. Technically, denoting the behaviour in the entanglement
entropy as a phase transition is inappropriate -- after all, the system
itself, \ie the state of the boundary field theory, does not change at
all. However, as we will see below, in our holographic calculation of
the entanglement entropy, there are competing saddle points and the
dominant saddle point shifts at a critical value of the width. Of
course, this behaviour is reminiscent of that seen in holographic
calculations describing thermodynamic phase transitions \cite{phasar}
and so we adopt the nomenclature `phase transition' to convey this
picture. The phase transition is first order and so the entanglement
entropy is continuous at the critical width $\l_t$, however, the
derivative $dS/d\l$ is discontinuous at this point. As a result, the
c-function drops discontinuously at the phase transition.

Implicitly, we are assuming that the holographic RG flows studied below
are solutions of Einstein gravity and hence the entanglement entropy is
determined by eq.~\reef{define}. Explicitly, our RG flow geometries
take the form given in eq.~\reef{metric0} and so are defined by giving
the conformal factor $A(r)$. Here we note that in all of the examples
we consider, $A''(r)\le0$ and so the geometry could solve Einstein's
equations with matter fields satisfying the null energy condition. In
appendix \ref{bulk}, we consider one approach to constructing an
appropriate scalar field theory that could realize the latter. In any
event with $A''(r)\le0$, the holographic c-theorem of section \ref{one}
will be satisfied. That is, $dc_d/dr_m\ge0$ or alternatively, the
c-function decreases monotonically as the corresponding extremal
surface extends deeper into the bulk geometry, as will be shown below.

In general, we will consider arbitrary values of the boundary dimension
in the following. However, to begin, we consider a very simple example
of a step flow and the discussion will be limited to the case $d=2$,
\ie a three-dimensional bulk. The step profile consists of two AdS
geometries with different curvature scales are patched together at some
finite radius. With such a simple profile, the behaviour of the
entanglement entropy and the c-function can be determined analytically.
Our analysis with $d=2$ is easily extended to higher $d$ but we do not
present the results here. In the subsequent subsection, we also examine
smooth profiles describing an holographic RG flow and allow for
arbitrary $d$. However, numerical analysis is required to understand
the behaviour of the entanglement entropy for these smooth profiles.

\subsection{Step profile} \label{step}

We limit the discussion here to three-dimensional gravity and consider
a bulk geometry\footnote{Various aspects of the flow of entanglement
entropy in this example was also studied in \cite{albash} for $d=2$, 3
and 4.} which patches together two AdS regions with different
curvatures at some finite radius $r=r_0$. Using the metric ansatz in
eq.~\eqref{metric0}, the conformal factor $A(r)$ is given by
 \be
A(r)=\bigg\{
 \begin{array}{l l l}
A_{IR}(r)=\frac{r\,-\,r_0}{L_{IR}}+\frac{r_0}{L_{UV}} & \textrm{for }
r\leq r_0 &  \\
A_{UV}(r)={r\ov L_{UV}} &\textrm{for }r\geq r_0 & \,,
\end{array}
 \labell{ex1x1}
 \ee
where $L_{UV}$ and $L_{IR}$ correspond to the AdS radius in the UV and
IR regions, which we denote as AdS$_{UV}$ and AdS$_{IR}$ in the
following. The constant term added to $A(r)$ in the IR region ensures
that the conformal factor is continuous at $r=r_0$. Of course, it is
not differentiable there and some stress energy with $\delta$-function
support would be required to make this geometry a solution of
Einstein's equations. As discussed in previous sections, there is a
conserved quantity \eqref{onex2} which plays an important role in
determining to the entanglement entropy and the c-function. Clearly,
there are two classes of minimal area surfaces in this geometry, namely
those that stay only in AdS$_{UV}$ and those that penetrate deep enough
into the bulk so that the minimal radius $r_m$ is in AdS$_{IR}$. In
either case, the conserved quantity $K_2$ is given by
eq.~\eqref{onex10} and hence we have $K_{UV}=e^{-A_{UV}(r_m)}$ for
$r_m\ge r_0$ and $K_{IR}=e^{-A_{IR}(r_m)}$ for $r_m\le r_0$. To
regulate the entanglement entropy, all of these surfaces are terminated
at a large cut-off radius $r=r_c$ in the UV region.

To find the minimal area surface, we will solve eq.~\eqref{onex2} for
$x=x(r)$. First, we can invert the latter equation to find
 \be
{dx \ov dr} \,=\,  {e^{-2A(r)} \ov \sqrt{K_2^2 -e^{-2A(r)}}}\,.
 \labell{nw1}
 \ee
Above we have discarded the root with an overall minus sign because we
will only consider the branch of solutions covering the interval $x
\in[0,\l/2]$ in the following, for which $dx/dr \ge0$. For the minimal
surfaces that stay entirely in AdS$_{UV}$, we can easily integrate
\eqref{nw1} to find
 \be
x \,=\, L_{UV} \sqrt{K_{UV}^2-e^{-2r/L_{UV}}}\,,
 \labell{nw2}
 \ee
using $A_{UV}$ as given in eq.~\eqref{ex1x1}. The integration constant
is chosen here so that $x=0$ at $r=r_m$, which also implies that
 \be
\ell = 2 L_{UV} K_{UV}=2L_{UV} e^{-r_m/L_{UV}}\,.
 \labell{bowm1}
 \ee
As noted above, this solution is valid for $r_m\ge r_0$, which implies
$\l \leq \l_{2}$ where
 \be
 \l_2\equiv 2L_{UV} e^{-r_0/L_{UV}}\,.
 \labell{bow0}
 \ee

Next we turn to the second class of minimal area surfaces, which
penetrate into AdS$_{IR}$. In this case, we have $K_2=K_{IR}=
e^{-A_{IR}(r_m)}$ with the turning point $r_m$ in AdS$_{IR}$. We divide
the relevant solutions of eq.~\reef{nw1} in two parts:
$x_{IR}(r)\in[0,x_t]$ describes the portion of the extremal surface in
AdS$_{IR}$ and $x_{UV}(r)\in[x_t,\l/2]$ represents the part in
AdS$_{UV}$. Here we have defined the transition point $x_t$ such that
$x_t=x_{IR}(r_0)=x_{UV}(r_0)$. Now integrating eq.~\eqref{nw1} with the
appropriate conformal factor \reef{ex1x1} for each segment, we find
 \bea
x_{IR} &\,=\,& L_{IR} \sqrt{K_{IR}^2-e^{-2A_{IR}(r)}} \nonumber \\
x_{UV} &\,=\,& L_{UV}\sqrt{K_{IR}^2-e^{-2A_{UV}(r)}} +{\l \ov 2} -
L_{UV} K_{IR} \,. \labell{nw3}
 \eea
Above, the integration constants were chosen so that $x_{IR}=0$ at
$r=r_m$ and $x_{UV}=\l/2$ as $r\to\infty$. Combining these solutions at
$x_{IR}(r_0)=x_{UV}(r_0)=x_t$, we find
 \bea
 x_t\,&=&\,{L_{IR} \ov
L_{UV}-L_{IR}} \left( L_{UV} K_{IR} -{\l \ov 2} \right)\,,
 \labell{nw4}\\
\l \,&=&\, 2 L_{UV}  K_{IR} - 2(L_{UV}-L_{IR})
\sqrt{K_{IR}^2-e^{-2r_0/L_{UV} }} \,.
 \labell{nw5}
 \eea
Here we have ensured that the solution \eqref{nw3} is continuous at
$r=r_0$ but the first derivative is also continuous at this transition
point because of the form of eq.~\reef{nw1} and the continuity of the
conformal factor. Implicitly, eq.~\reef{nw5} gives the relation between
$\l$ and $r_m$ since $K_{IR}= e^{-A_{IR}(r_m)}$. As the physically
relevant quantity in the boundary theory is $\l$, we invert this
relation to find the following two solutions for $K_{IR}$:
 \be
K_{IR\pm} \,=\,{L_{UV} \l \pm (L_{UV}-L_{IR})\sqrt{\l^2 - 4 L_{IR}
(2L_{UV}-L_{IR} ) e^{-2r_0/L_{UV}} } \ov 2 L_{IR}(2L_{UV}-L_{IR})} \,.
 \labell{nw6}
 \ee
Above, both of these roots provide real solutions for $\l\ge\l_1$ with
 \be
\l_1\equiv 2 \sqrt{L_{IR}(2L_{UV}-L_{IR})}\, e^{-r_0/L_{UV}}\,.
 \labell{bow2}
 \ee
It is also useful to define $r_1$, the value of the minimum radius at
$\l=\l_1$, for which we find
 \be
e^{-r_1/L_{IR}}\equiv\frac{L_{UV}}{\sqrt{L_{IR}(2L_{UV}-L_{IR})}}\,
e^{-r_0/L_{IR}}\,.
 \labell{bow}
 \ee
Now the root $K_{IR+}$ is a monotonically increasing function of $\l$
over the range $\l_1\le\l\le\infty$ and for any $\l$ in this range,
there is a consistent solution for the extremal surface. The
corresponding values of the minimal radius are $r_1\ge r_m\ge-\infty$.
Now the second root $K_{IR-}$ decreases for $\l\sim\l_1$, however, it
is an increasing function for large values of $\l$. $K_{IR-}$ has a
single minimum at $\l=\l_2$, \ie precisely the width defined in
eq.~\reef{bow0} for the discussion of solutions remaining entirely in
AdS$_{UV}$. At this minimum, $K_{IR-}$ takes the value
$e^{-A_{IR}(r_0)} = e^{-r_0/L_{UV}}$. We find that $K_{IR-}$ yields a
consistent solution for the extremal surface as long as
$\l_1\le\l\le\l_2$. However, for $\l>\l_2$, the solutions corresponding
to $K_{IR-}$ are inconsistent, \eg $dx/dr$ is not positive throughout
the range $x\in[0,\ell/2]$. We note that for the consistent solutions,
while $\l$ runs from $\l_1$ to $\l_2$, the minimum radius of these
surfaces $r_m$ increases from $r_1$ to $r_0$. That is, in contrast to
the previous two families of solutions, here we have $d\l/dr_m>0$!

Hence the following picture has emerged for the extremal surfaces:
Beginning with the minimal radius in the range $\infty>r_m\ge r_0$,
there is a family of extremal solutions which remain entirely in
AdS$_{UV}$. As can be seen from eq.~\reef{bowm1}, the width $\l$
increases monotonically as $r_m$ decreases, reaching the maximum
$\l=\l_2$ when $r_m=r_0$. Below this point, we make a transition to a
new family of solutions which begin to penetrate into AdS$_{IR}$. For
$r_0\ge r_m\ge r_1$, the relevant family of extremal surfaces
corresponds to the branch with $K_{IR-}$. In this regime, $\l$ actually
decreases as $r_m$ continues to decrease, \ie $d\l/dr_m>0$. When $r_m$
reaches $r_1$, as given in eq.~\reef{bow}, $\l=\l_1$ and we make
another transition to the third family of extremal surfaces. These
solutions correspond to the branch with $K_{IR+}$. In this regime
$r_1>r_m>-\infty$, $\ell$ again increases monotonically as $r_m$
decreases. Figure \ref{lvsr} illustrates this behaviour for all three
families of extremal surfaces. Now for any particular value of the
turning point radius $r_m$, we see there is unique extremal surface.
However, if we consider the solutions as a function of the strip width
$\ell$, there is a unique solution for $\l<\l_1$ and $\l>\l_2$. In the
intermediate range $\l_1\le\l\le\l_2$, there are in fact three possible
extremal surfaces for any given width. Given three possible saddle
points, we are instructed in eq.~\reef{define} to find the extremal
surface with the minimum area in order to evaluate the entanglement
entropy. This situation with multiple saddle points is also the typical
scenario that one encounters in the holographic description of a
thermodynamic phase transition \cite{phasar} and in fact, we will find
the latter extends to the present situation. That is, we see below that
the entanglement entropy undergoes a `first order phase transition'.
\FIGURE{
\includegraphics[width=0.9\textwidth]{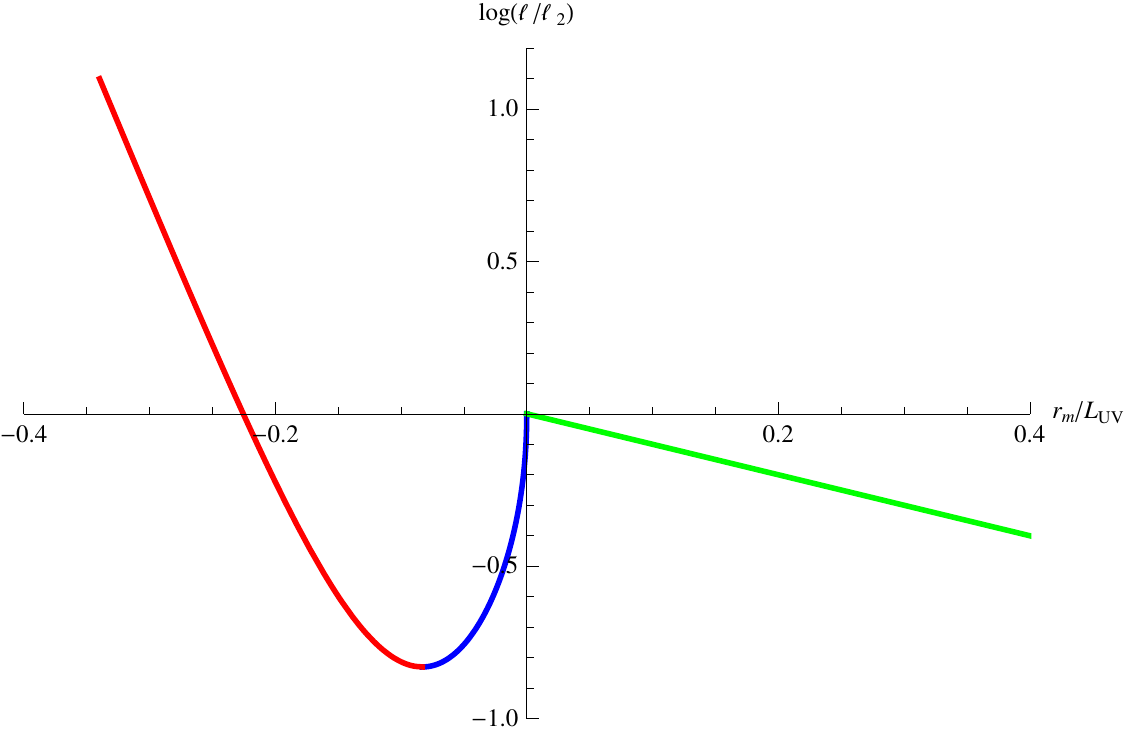} %
\caption{(Colour Online) This plot illustrates the behaviour of $\l$ as
a function of $r_m$ for all three families of extremal surfaces. The
red, blue and green portions of the curve correspond to the
contributions coming from $S_{IR}(K_{IR+})$, $S_{IR}(K_{IR-})$ and
$S_{UV}$, respectively. In particular, we see that $d\l/dr_m\le0$ for
$S_{IR}(K_{IR+})$ and $S_{UV}$, while  $d\l/dr_m\ge 0$ for
$S_{IR}(K_{IR-})$. For this plot, we chose $L_{UV}=1$, $L_{IR}=0.1$ and
$r_0=0$.} \label{lvsr}}

Hence having found the solutions for the extremal surfaces, we will
present an entropy, \ie $S=2\pi A(v)/\lp$, for each of these surfaces.
But, of course, in the intermediate regime described above, the true
entanglement entropy is given by the solution which minimizes this
quantity. Let us begin with the solutions \eqref{nw2} which remain
entirely in AdS$_{UV}$. For this case, the entropy turns out to be
 \be
S_{UV}\,=\, {4 \pi L_{UV} \ov \lp}\, \log\left({\l \ov \delta}\right)
+\mathcal{O}(\delta) \,.
 \labell{nw7}
 \ee
%rcm check if O(delta) term really is there
Here the result is expressed in terms of a short-distance cut-off in
the boundary theory $\delta$, which is related to the radial cut-off
by\footnote{This matches the standard cut-off $z_{min}=\delta$ in
Poincar\'e coordinates where the AdS metric takes the form
$ds^2=(L_{UV}^2/z^2)\left(\eta_{ij}dx^idx^j +dz^2\right)$ -- \eg see
\cite{casini9}.}
 \be
r_c=L_{UV}\,\log\left(L_{UV}/\delta\right)\,.
 \labell{cow0}
 \ee
Now let us consider the extremal surfaces given by eqs.~\eqref{nw3} to
\eqref{nw6}, which penetrate into AdS$_{IR}$. For this case, the
calculation of entanglement entropy results
 \begin{align}
S_{IR}\,&=\,{4 \pi \ov \lp} \int_0^{x_t} dx \, \sqrt{e^{2 A_{IR}(r)}+
\dot{r}^2} + {4 \pi \ov \lp} \int_{x_t}^{(\l-\epsilon)/2} dx \, \sqrt{e^{2
A_{UV}(r)}+ \dot{r}^2} \labell{nw9} \\
&=\, \frac{2 \pi L_{IR} }{\lp} \log\left[\frac{ L_{IR} K_{IR} + x_t}{L_{IR}
K_{IR} -x_t}\right] + \frac{2 \pi L_{UV}}{\lp} \log\left[
\frac{4L_{UV} K_{IR}-\epsilon}{4 L_{UV} K_{IR} - \l + 2 x_t}
\,\frac{\l - 2 x_t}{\epsilon}\right] \,.
\nonumber
%
%&=\, \frac{2 \pi L_{IR} }{\lp} \log\left[\frac{ L_{IR} K_{IR} + x_t}{L_{IR}
%K_{IR} -x_t}\right] + \frac{2 \pi L_{UV}}{\lp} \log\left[\frac{4L_{UV} K_{IR}-\epsilon}{\epsilon}
%\right]-\frac{2 \pi L_{UV}}{\lp} \log\left[\frac{4 L_{UV} K_{IR} - \l + 2 x_t}{\l - 2 x_t}\right] \,.
%\nonumber
\end{align}
We convert the cut-off $\epsilon$ above to $\delta$ using
$r(x=(\l-\epsilon)/2,\l)=r_c$, which yields
 \be
\epsilon \,\simeq\, {\delta^2 \ov  L_{UV} K_{IR}} \,.
 \labell{nw10}
 \ee
Using this relation, we can write
 \bea
S_{IR} &\,=\,& \frac{4 \pi L_{UV}}{\lp} \log\left[\frac{2L_{UV}
K_{IR}}{\delta}\right] +\frac{2 \pi L_{IR} }{\lp} \log\left[\frac{
L_{IR} K_{IR} + x_t}{ L_{IR} K_{IR} -x_t}\right] \nonumber \\
&& \qquad \qquad \qquad \qquad \qquad -\frac{2 \pi L_{UV}}{\lp}
\log\left[\frac{4 L_{UV} K_{IR} - \l + 2 x_t}{\l - 2
x_t}\right] \,.
 \labell{nw11}
 \eea
This result is valid for both of the roots $K_{IR\pm}$ given in
eq.~\reef{nw6}.

Now as described above, the UV family of solutions \eqref{nw2} provide
the unique extremal surface for any $\l<\l_1$ and hence the
entanglement entropy is given by $S=S_{UV}$ in this regime. Similarly,
for $\l>\l_2$, the extremal surface is again unique and hence the
entanglement entropy is given by $S=S_{IR}(K_{IR+})$. In the
intermediate regime $\l_1\le\l\le\l_2$, we have three extremal surfaces
and we must identify which of these yields the minimal entropy. In
particular, we always find $S_{IR}(K_{IR+})\le S_{IR}(K_{IR-})$ and
hence the branch with $K_{IR-}$ never plays a role in determining the
physical entanglement entropy. Therefore the latter is found by
comparing $S_{UV}$ and $S_{IR}(K_{IR+})$. It turns out that these two
entropies are equal for some critical width $\l_t$ with
$\l_1\le\l_t\le\l_2$. Further $S_{UV} > S_{IR}(K_{IR+})$ for $\l>\l_t$
and $S_{UV} < S_{IR}(K_{IR+})$ for $\l<\l_t$. Hence we find that, the
entanglement entropy for the step profile \eqref{ex1x1} is given by
 \be
S \,=\,\left\{
 \begin{array}{l l}
S_{UV}(\l) & \textrm{for }\l \leq \l_t  \\
S_{IR}(K_{IR+}(\l)) &\textrm{for } \l \geq \l_t  \,,
\end{array} \right.
\labell{nw11x1}
 \ee
with $\l_t$ defined by $S_{UV}(\l_t) =S_{IR}(K_{IR+}(\l_t))$. In
particular, we observe that the entanglement entropy exhibits a first
order phase transition at the critical width $\l=\l_t$. Further, we
note that at transition, $S(\l)$ is continuous but not differentiable.
We have illustrated all of this behaviour in figure \ref{2ee}, which
plots $S(\l)-S(\l_2)$ versus $\log(\l/\l_2)$ for specific values of the
parameters, $L_{UV}$, $L_{IR}$ and $r_0$, defining the
profile.\footnote{Since $S(\l)$ diverges as $\delta\to0$, we plot the
difference $S(\l)-S(\l_2)$ which is independent of $\delta$.}
\FIGURE{
\includegraphics[width=.9\textwidth]{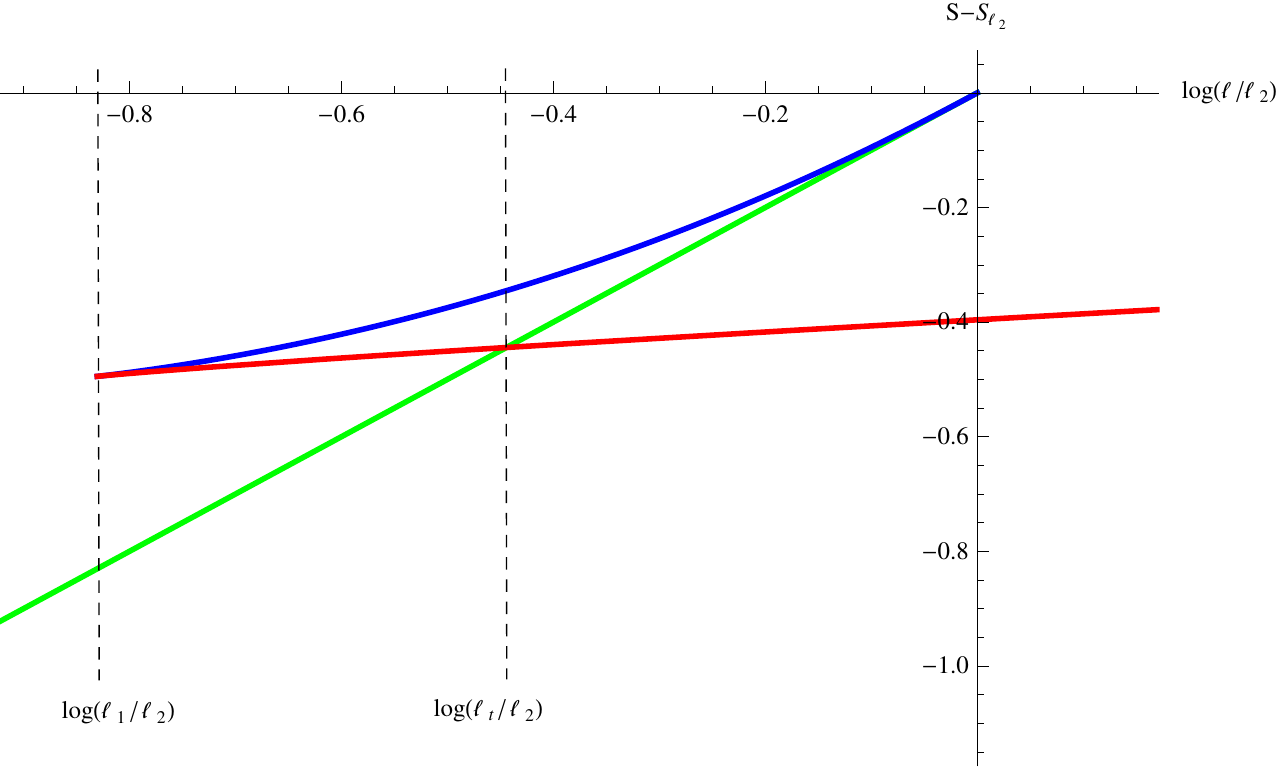}
\caption{(Colour Online) This plot illustrates the entropy for all
three families of extremal surfaces as a function of the width of the
strip $\l$. The green curve corresponds to $S_{UV}$; the red, to
$S_{IR}(K_{IR-+})$; and the blue, to $S_{IR}(K_{IR-})$. The phase
transition occurs at $\l=\l_t$ where the red and green curves cross.
For this plot, we have chosen $L_{UV}=1$, $L_{IR}=0.5$ and $r_0=0$.}
\label{2ee}}

Given the entanglement entropy \reef{nw11x1}, we turn to the
calculation of the c-function defined in eq.~\reef{dim2}. We could
proceed here by explicitly differentiating the various expressions
above, \eg eq.~\eqref{nw11}, with respect to $\l$ to determine $c_2$.
However, this calculation only verifies the final result which was
already determined in our general analysis in section \ref{one}, namely
eq.~\reef{onex9}. In fact, this result applies for all three families
of extremal surfaces and so we have
 \be
{c}_2 \,=\,\left\{
 \begin{array}{l l}
{6 \pi \ov \lp}{\l \ov K_{UV}} & \textrm{from }S_{UV}  \\
{6 \pi \ov \lp}{\l \ov K_{IR-}} &\textrm{from }S_{IR}(K_{IR-}) \\
{6 \pi \ov \lp}{\l \ov K_{IR+}} &\textrm{from }S_{IR}(K_{IR+})  \,,
\end{array} \right.
\labell{nw14}
 \ee
where $K_{UV}=\l/(2L_{UV})$ and $K_{IR\pm}$ are given in
eq.~\eqref{nw6}. Figure \ref{fig1}a plots $c_2$ as a function of the
turning point radius -- or rather $e^{r_m/L_{UV}}$. We see that
$d{c}_2/dr_m \geq 0$ everywhere in the figure, which is again in
keeping with the expectations of our general analysis in section
\ref{one}. However, because of the phase transition, not all values of
$r_m$ are relevant for the c-function \reef{dim2} evaluated on the
physical entanglement entropy \reef{nw11x1}.  In figure \ref{fig1}a,
the region between the vertical dashed lines is excluded and the
physical c-function jumps discontinuously between the values at the
points labeled $A$ and $B$. This behaviour is also illustrated in
figure \ref{fig1}b where the c-function in eq.~\reef{nw14} is plotted
as a function of the ratio $\l/\l_t$. The phase transition at $\l=\l_t$
again takes $c_2$ between the points labeled $A$ and $B$, which now lie
on the same vertical dashed line in this figure. That is, at this
critical value of the strip width, the c-function drops from the value
given by $S_{UV}$ (on the green curve) to that given by
$S_{IR}(K_{IR+})$ (on the red curve). Again this discontinuity arises
because of the first order nature of the phase transition, \ie the
entanglement entropy is continuous but not differentiable at this
point. If we consider only the physical values of $c_2$, then we also
find $d{c}_2/d\l \leq 0$ in keeping with the general expectations of
field theory analysis of \cite{casini2}. Of course, figure \ref{fig1}b
also illustrates that $d{c}_2/d\l> 0$ on the branch associated with
$K_{IR-}$. However, as emphasized above, this family of saddle points
is not relevant of the physical entanglement entropy \reef{nw11x1}. The
`unusual' behaviour of the c-function on this branch arises because
$d\l/dr_m \geq 0$ for this family of solutions. Given the behaviour
illustrated by this simple example, it seems likely that in general any
branch of extremal surfaces with the latter property will correspond to
unstable saddle points which are not physically relevant.
\FIGURE[!ht]{
\begin{tabular}{cc}
\includegraphics[width=0.5\textwidth]{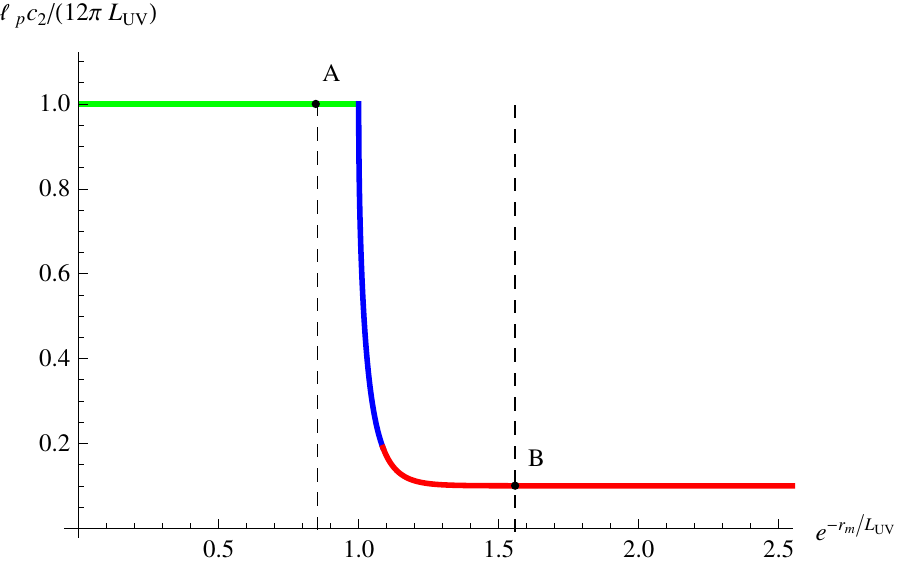}&
\includegraphics[width=0.5\textwidth]{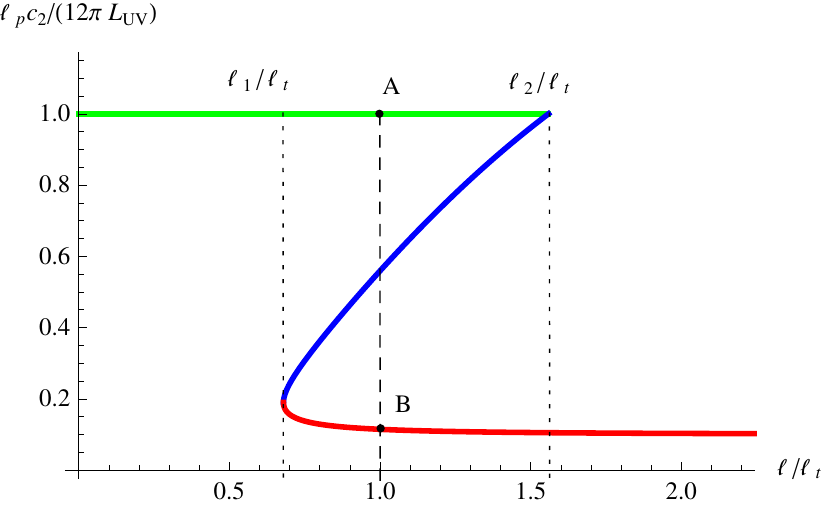}\\
(a) & (b)
\end{tabular}
\caption{(Colour Online)  Panel (a) plots the c-function \eqref{nw14}
as a function of $e^{-r_m/L_{UV}}$. This plot clearly illustrates that
$c_2$ decreases monotonically as $r_m$ decreases, in accord with the
analysis of section \ref{one}. However, the region between the vertical
dashed lines is excluded by the phase transition in the entanglement
entropy. Rather the physical c-function jumps from the point labeled A
to that labeled B. Panel (b) plots the c-function with respect to
$\l/\l_t$. One can see that $c_2$ is multi-valued in the region
$\l_1\le\l\le\l_2$, as noted in eq.~\reef{nw14}. Again at $\l=\l_t$,
the c-function drops discontinuously between the points labeled A and
B. The red, blue and green portions of both curves correspond to the
contributions coming from $S_{IR}(K_{IR+})$, $S_{IR}(K_{IR-})$ and
$S_{UV}$, respectively. For this plot, we chose $L_{UV}=1$,
$L_{IR}=0.1$ and $r_0=0$.} \label{fig1}}

\subsection{Smooth profiles} \label{smooth}

The simple example in the previous section has alerted us to the
possibility that the entanglement entropy $S(\l)$ may experience a
phase transition with respect to changing the strip width $\l$.
However, one should worry that this result is an artifact of the
artificial shape of the step profile in eq.~\reef{ex1x1}. Hence we
consider some smooth profiles in this section and examine to what
extent this phase transition survives for these more realistic
holographic RG flows. Again our definition of the holographic
entanglement entropy is given in eq.~\reef{define} and so implicitly we
are assuming that the bulk geometry is a solution of Einstein's
equations. In appendix \ref{bulk}, we consider the scalar field theory
that would be necessary to realize the latter. In this section, we will
consider arbitrary values of the boundary dimension $d$.

Let us first consider a smooth flow between the UV and IR fixed points
with the following conformal factor
 \be
e^{A(r)}\,=\,e^{r/ L} \,\big(2\cosh\!\left({r/
R}\right)\big)^{-\gamma}\,,
 \labell{nw15}
 \ee
Notice that $A(r)\simeq r/L - \gamma r/R$ in the limit $r\to + \infty$
and $A(r)\simeq r/L + \gamma r/R$ for $r\to - \infty$. Hence the
geometry approaches AdS space in both of these limits with
 \be
{1\ov L_{UV}}\,\equiv\, {1\ov L}-\frac\gamma{R} \quad \textrm{and}
\quad {1\ov L_{IR}}\,\equiv\, {1\ov L} +\frac\gamma{R} \,.
 \labell{nw16}
 \ee
The parameter $R$ controls the sharpness of the transition in the
holographic flow between the UV and IR fixed points while the change in
the AdS scale is controlled by the combination $\gamma/R$. In the limit
$R\to 0$ with $\gamma/R$ fixed, we would recover a step profile of the
form given in eq.~\eqref{ex1x1}.

To proceed further in examining the possibility of a phase transition,
we used the analysis presented in previous sections and examined the
extremal surfaces numerically for the above holographic flow profile.
First, using eq.~\reef{twox3a}, the equation determining the shape of
the extremal surfaces is reduced to a first order equation,
 \be
{dx \ov dr} \,=\,  {e^{-dA(r)} \ov \sqrt{K_d^2 -e^{-2(d-1)A(r)}}}\,.
 \labell{gamma0}
 \ee
as appears in eq.~\reef{nw1} for $d=2$. Then families of surfaces are
easily constructed as a function of the turning point radius $r_m$
using eq.~\reef{onex10a}. Numerically integrating from the turning
point out to the asymptotic region, we can then determine $\l(r_m)$.

Let us add a few more details about the numerical analysis: Near
$r=r_m$, one finds that $x\sim \sqrt{r-r_m}$  and hence
eq.~\eqref{gamma0} is singular precisely at $r=r_m$, the putative
starting point of our numerical integration. So to simplify the
numerical analysis, we define
 \be y(r)=\sqrt{r-r_m}\, x(r)\,, \labell{newco} \ee
for which the equation of motion \eqref{gamma0} becomes
 \be
{dy\ov dr}\,=\, {y \ov 2(r-r_m)}+ {\sqrt{r-r_m} \, e^{-dA(r)} \ov
\sqrt{K_d^2-e^{-2(d-1)A(r)}} } \,.
 \labell{neweom}
 \ee
With this new coordinate, $y\sim (r-r_m)$ near $r=r_m$ and the right
hand side of the equation of motion \reef{neweom} is finite. To set the
initial conditions, we use eq.~\eqref{neweom} to find the leading terms
in a series expansion of $y(r)$ in $r-r_m$. Now we can numerically
integrate eq.~\eqref{neweom} out from the turning point $r=r_m$ to
large asymptotic values of $r$ and find the the strip width $\l$ using
the relation
 \be
\l\,=\, \lim_{r\to \infty} {2\,y(r) \ov \sqrt{r-r_m}}\,.
 \labell{nothing9}
 \ee

Now as discussed previously, the appearance of a phase transition is
directly related to the appearance of a regime where $d\l/dr_m>0$.
Figure \ref{lvsrmd2} illustrates that such behaviour still arises for a
range of parameters in the smooth profile \reef{nw15}. However, as
shown in the figure when $R$ grows (holding $\gamma/R$ fixed), this
region decreases and eventually $d\l/dr_m<0$ for all values of $r_m$.
That is, there exists a critical value $R_c$ such that, for $R<R_c$
there is a first order phase transition while for $R>R_c$, we observe a
smooth cross-over. At precisely $R=R_c$, there is a single point where
$d\l/dr_m=0$ and the slope is otherwise negative. In this case, the
phase transition would be second order.
\FIGURE[!ht]{
\includegraphics[width=0.8\textwidth]{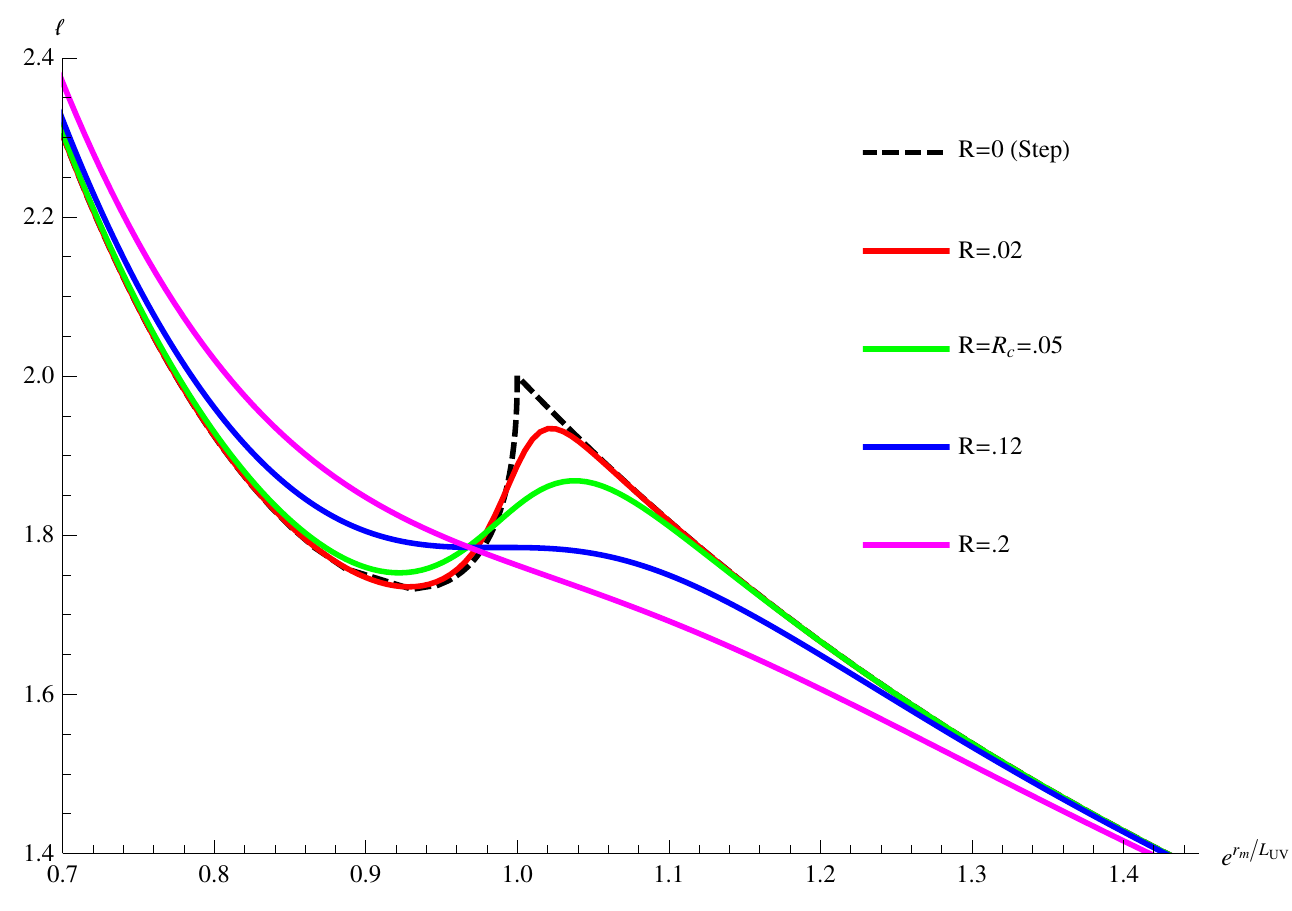}
\caption{(Colour Online) Plot of strip width $\l$ as a function of
$e^{r_m/L_{UV}}$ for various values of the profile width $R$. With
small $R$, there is a regime where $d\l/dr_m>0$. However, for large
$R$, $d\l/dr_m<0$ everywhere. There is a critical value $R_c$ for which
$d\l/dr_m$ reaches zero at a single point and is otherwise negative.
The plot was prepared using the smooth profile \eqref{nw15} with
$L=0.66$ and $\gamma=0.5 R$, as well as $d=2$. These parameter values
yield $L_{IR}=0.5$ and $L_{UV}=1$.} \label{lvsrmd2}}

Of course, one can go beyond the above analysis to identify the precise
point where the phase transition occurs given specific values of the
parameters in eq.~\reef{nw15} which produce a regime where
$d\l/dr_m>0$. Hence for some range of the width $\l$, there will be
multiple surfaces which locally extremize the entropy functional
\reef{twox1}. Determining which surface provides the dominant saddle
point requires carefully regulating the entanglement entropy and
comparing the values of finite parts of entropy for the competing
saddle points. This analysis is essentially the same as in section
\ref{step}, however, in the present case with a smooth conformal
factor, the profile $x(r)$ and the entropy integral are evaluated
numerically. We will not present any of these results here.
%\comment{add comments about finding phase transition point in thesis}

Let us now turn to the asymptotic expansion of the profile \reef{nw15}
and compare it to the Fefferman-Graham (FG) expansion given in
eqs.~\eqref{feffer} and \eqref{feffer1}. We find that
 \be
f(r)\,=\,\left(1+e^{-2r/R}\right)^{-\gamma} \simeq 1 - \gamma e^{-2
r/R}+\dots \,.
 \labell{nw17}
 \ee
which yields $\alpha=L_{UV}/R$ in eq.~\eqref{feffer1}. Now as described
in section \ref{review}, the natural holographic interpretation of this
flow would be that the UV fixed point is perturbed by a relevant
operator, which would be dual to by a scalar field in the bulk theory.
However, as discussed in appendix \ref{bulk}, applying this
interpretation to the bulk solution yields an upper bound on the
parameter $\alpha$ appearing in the FG expansion \reef{feffer1}, \ie
$\alpha \leq {d\ov 2}+1$. This bound then becomes a constraint on $R$,
the width of the holographic profile \reef{nw15}. That is,
 \be
R \geq {2 L_{UV}\ov d+2}\,.
 \labell{bfbound}
 \ee

Previously we found that the smooth holographic RG flows described by
eq.~\reef{nw15} will still yield a first order phase transition in the
entanglement entropy provided the width is sufficiently small, \ie
$R\le R_c$. Hence the lower bound given in eq.~\reef{bfbound} creates a
certain tension. Namely, if $R_c$ does not satisfy this lower bound, it
seems likely that the phase transition is still an artifact of the
artificial shape of the profile in eq.~\reef{nw15}. We examine this
question in figure \ref{rcplot}a, where we have plotted $(d+2)
R_c/2L_{UV}$ for different values of $L_{UV}/L_{IR}$. The bound
\reef{bfbound} implies that $(d+2) R/2L_{UV}>1$ and as the figure
illustrates, the latter can only be satisfied for sufficiently large
$d$, \ie $d \geq 6$. Hence the possibility of a phase transition is
called into question for the physical dimensions $d=2,3,4$.
\FIGURE[!ht]{
\begin{tabular}{cc}
\includegraphics[width=0.5\textwidth]{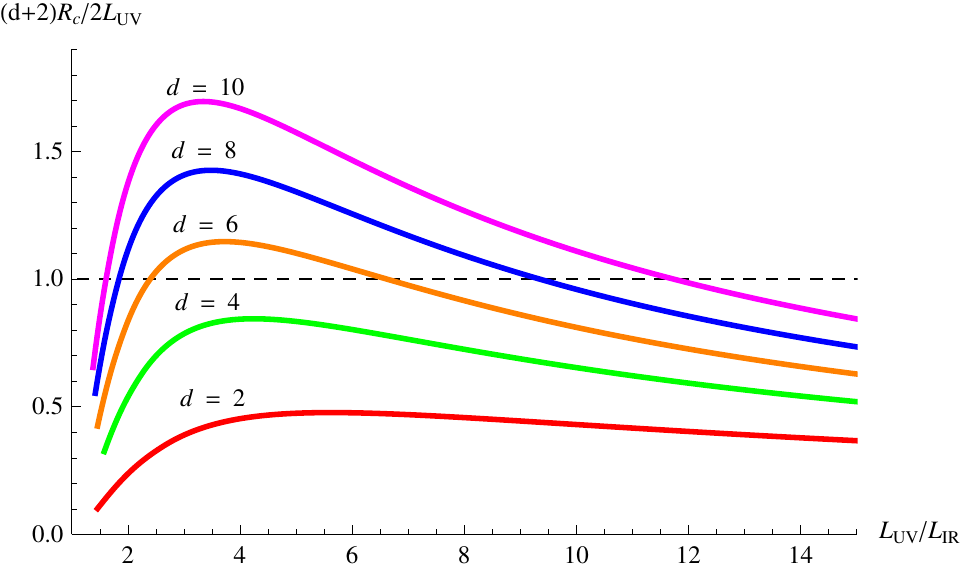}&
\includegraphics[width=0.5\textwidth]{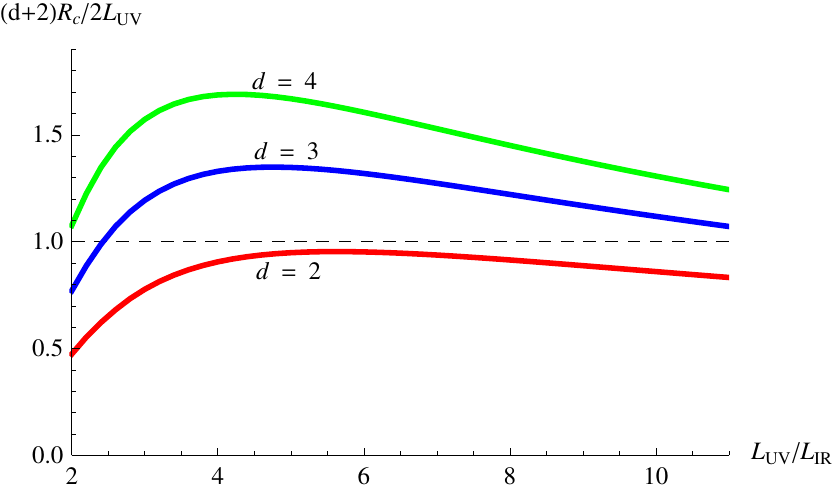}\\
(a) & (b)
\end{tabular}
\caption{(Colour Online) Panel (a) plots ${(d+2) R_c \ov 2 L_{UV}}$ as
a function of ${L_{UV}\ov L_{IR}}$ for various dimensions with the
two-parameter conformal factor \eqref{nw15}. Eq.~\reef{bfbound} implies
${(d+2) R_c \ov 2 L_{UV}}>1$ for a consistent interpretation of the
holographic RG flow. Panel (b) plots ${(d+2) R_c \ov 2 L_{UV}}$ as a
function of ${L_{UV}\ov L_{IR}}$ for the three-parameter conformal
factor \eqref{nw18} with $\sigma=5\times 10^{-4}$.} \label{rcplot}}

Now the profile \reef{nw15} was constructed to give a simple example
which would smooth out the step potential studied in the previous
subsection. One can easily generalize this profile to include more
independent parameters and study the effect on the phase transition.
Hence as another simple example, we consider the following conformal
factor
 \be
e^{A(r)}\,=\,e^{r/ L} \left(e^{2r/R}+2 \sigma + e^{-2r/R}
\right)^{-\gamma/2}\,.
 \labell{nw18}
 \ee
Of course, if one chooses $\sigma=1$, this profile reduces to the
previous one in eq.~\reef{nw15}.\footnote{With the choice $\sigma=0$,
eq.~\reef{nw18} also reduces to eq.~\reef{nw15} upon substituting
$R\to2R$ and $\gamma\to2\gamma$.} Hence the AdS scales in the UV and IR
limits are again given by eq.~\reef{nw16} with the new expression.
However, in this case, the sharpness of the transition between the
asymptotic UV and IR geometries is effectively controlled by both $R$
and $\sigma$. Given this new profile, we can readily extend the
previous analysis to find in which parameter regime $(R,\sigma)$ the
entanglement entropy undergoes a phase transition. We do not present
any details but a qualitative observation is that $R_c$ becomes larger
with smaller values of $\sigma$. Now considering the FG expansion in
this case, the metric function in eq.~\eqref{feffer1} becomes
 \be
f(r)\,=\,\left(1+2\sigma e^{-2r/R} + e^{-4r/R}\right)^{-\gamma/2} = 1 -
\gamma\,\sigma\, e^{-2 r/R}+\dots \,.
 \labell{nw17a}
 \ee
Hence we still have $\alpha=L_{UV}/R$ and the lower bound in
eq.~\reef{bfbound} remains unchanged. Hence, as shown in figure
\ref{rcplot}b, we find that $R_c$ can now satisfy this bound for $d
\geq 3$. Our expectation is that by further embellishing the form of
the holographic RG flow profile, we can also find realistic geometries,
\ie geometries satisfying eq.~\reef{bfbound}, which produce a phase
transition in the entanglement entropy for $d=2$ as well.

%%%%%%%%%%%%%%%%%%%%%%%%%%%%%%%%%%%%%%%%%%%%%%%%%%%%%%%%%%%%%%
\section{Holographic flow of c-function with GB gravity}\label{three}

In this section, we return to examining the general behaviour of the
c-function \eqref{candid} in holographic RG flows but now where the
bulk geometries are solutions of GB gravity
\reef{GBAction}.\footnote{We refer the interested reader to appendix
\ref{gravity} for a brief discussion describing the explicit
construction of such solutions.}  As noted in section \ref{holoEE}, the
calculation of holographic entanglement entropy in GB gravity requires
that we extremize the entropy functional given in eq.~\eqref{gbstrip}
\cite{ent1,ent2}. Evaluating this functional for the strip geometry
yields the expression in eq.~\reef{functional}. The extremal surfaces
are again characterized by a conserved quantity which now takes the
form given in eq.~\eqref{conserved}. The latter is most simply
evaluated by considering the turning point of the extremal surface
where $r=r_m$ and $\dr=0$ so that eq.~\eqref{conserved} yields
 \be
K_d(r_m) \,=\, e^{-(d-1)A(r_m)}\,, \labell{threex1}
 \ee
precisely as was found before for Einstein gravity.

Recall that with Einstein gravity, we found a simple relation between
$K_d$ and $dS/d\l$ -- see eq.~\eqref{twox14}. In the following, we will
show that this same relation extends to GB gravity. To simplify the
discussion, we denote the integrand in eq.~\reef{functional} as
$\mathcal{L}$ and then $K_d$ can be expressed as
 \be
{1 \ov K_d(\l)}\,=\, \mathcal{L} - {d \mathcal{L} \ov d \dr} \dr \,.
 \labell{threex2}
 \ee
Now we vary the entanglement entropy functional \eqref{functional} with
respect to the width of the strip $\l$ to find
 \be
{d S_{GB}\ov d \l }  \,=\,{4 \pi H^{d-2} \ov \lp^{d-1}} \left[ {1\ov
2}\left(1-{d\epsilon \ov d\l} \ri) \mathcal{L} \Big|_{x={\l-\epsilon\ov
2}} +\int_0^{{\l-\epsilon \ov 2}} dx \left( {\delta\mathcal{L} \ov
\delta r}\, {\partial r\ov\partial\l} + {\delta \mathcal{L} \ov \delta
\dr} \,{\partial \dr \ov
\partial \l}  \ri)   \ri]\,.
 \labell{threex3}
 \ee
Note that there is an extra overall factor of $2$ above to since we are
only integrating over half of the bulk surface, \ie from the turning
point $(x,r)=(0,r_m)$ to the boundary $(x,r)=((\l-\epsilon)/2,r_c)$.
Now surfaces extremizing eq.~\reef{functional} will satisfy
 \be
{\delta \mathcal{L} \ov \delta r} ={\partial \ov \partial x} \left(
{\delta \mathcal{L} \ov \delta \dr} \ri) \,.
 \ee
Further eq.~\reef{onex4} still applies in the present analysis and so
allows us to express $(1-d\epsilon/d\l)$ in terms of derivatives of the
profile $r(x,\l)$. With both of these expressions, we are able to
simplify eq.~\eqref{threex3} to take the form
 \bea
{d S_{GB}\ov d \l }  &\,=\,& {4 \pi H^{d-2} \ov \lp^{d-1}} \left[
-{\mathcal{L} \ov \dr}\,{\partial r\ov\partial\l}\Big|_{x={\l-\epsilon
\ov 2}} + \left[ {d\mathcal{L} \ov d \dr} {\partial r\ov\partial\l}
\ri]_{x=0}^{{\l-\epsilon \ov
2}} \ri] \nonumber \\
%
%&\,=\,& {4 \pi H^{d-2} \ov \lp^{d-1}} \bigg[ -{1 \ov K_d} {r' \ov \dr}
%\bigg|_{{\l-\epsilon \ov 2}} -  {d \mathcal{L} \ov d \dr}r'  \bigg|_{0} \bigg]
%\nonumber \\
%
&\,=\,& - {4 \pi H^{d-2} \ov \lp^{d-1}K_d(\l)} {1 \ov \dr}\, {\partial
r\ov\partial\l }\bigg|_{x={\l-\epsilon \ov 2}} \,.
 \labell{threex4}
 \eea
Here the boundary term at $x=0$ vanishes because $\dr=0$ there.
Eq.~\eqref{threex2} was then used to simplify the remaining terms with
$K_d(\l)$. We note that precisely the same expression as above appeared
in the analysis of $d S/d \l$ with Einstein gravity -- see
eq.~\reef{twox7}.

The next step is to show that the ratio ${1\ov \dr}{\partial r\ov
\partial \l}$ has the same simple boundary limit as found previously
in eq.~\reef{twox13} with Einstein gravity. The analogous procedure
would call for solving eq.~\eqref{conserved} to find $\dr$. However, in
the present case, we would find a cubic equation in $\dr^2$ and which
would in general have six distinct roots. The relevant root would be
that which in the limit $\lambda \to 0$ is continuously connected to
the solution \reef{twox11b} found with Einstein gravity. While it is
possible to carry out this procedure analytically, it is not very
illuminating. Rather we note that we are interested in the behaviour
near the asymptotic boundary where the geometry approaches AdS space
and the conformal factor takes the form $A(r)=r/\tilde{L}$. Now it is
sufficient to expand near the boundary where $e^{A}$ is very large and
the leading contribution to $\dr$ becomes
 \be
\dr \,\simeq\, e^{dr/\tL}\,K_d \, \left(1-2\lambda \fin \right)\, \,.
\labell{threex12}
 \ee
This equation is easily solved to yield $x(r)$ near the boundary. Of
course, the integration constant is chosen to satisfy the boundary
condition $x(r\to\infty)\to\l/2$. Using this asymptotic solution for
the profile of the extremal surface, it is easy to confirm that
 \be
{1 \ov \dr} {\partial r \ov \partial \l} \bigg|_{x=\l/2}\,=\,-{1\ov
2}\,,
 \labell{threex13}
 \ee
as desired. Note that this relation is independent of the GB coupling
$\lambda$.

Substituting eq.~\eqref{threex13} into eq.~\eqref{threex4}, we arrive
at
 \be
{d S_{GB} \ov d \l } \,=\,{2 \pi H^{d-2} \ov \lp^{d-1}} {1\ov K_d(\l)}
\,, \labell{threex14}
 \ee
which precisely matches the expression \reef{twox14} found with
Einstein gravity. It seems that this result is quite general. The first
key ingredient is, of course, the conserved quantity \reef{threex2}.
The other necessary ingredient is that the asymptotic geometry
approaches AdS space, which seems sufficient to produce the simple
expression in eq.~\reef{threex13}. Hence we expect that the expression
\eqref{threex14} should be general to all cases with these two basic
features.

Next, we turn to the flow of c-function \eqref{candid} as we change the
minimum radius $r_m$ of the extremal surfaces, as considered for
Einstein gravity in section \ref{one}. Given eq.~\eqref{threex14}, our
starting point for the c-function \reef{candid} is precisely the same
as in eq.~\reef{rex1}, \ie
 \be
{c}_d \,=\, {2 \, \pi \,\beta_d \ov \lp^{d-1}} { \l^{d-1} \ov K_d(\l)}
\,,
 \labell{rex1gb}
 \ee
Hence using eq.~\eqref{threex1}, we find
 \be
{d {c}_d \ov d r_m } \,=\, {2 \, \pi (d-1) \beta_d \l^{d-2} \ov
\lp^{d-1} K_d } \left({d \l \ov d r_m} +A'(r_m) \, \l  \right)\,.
 \labell{threex15}
 \ee
Following our previous analysis, we express $\l$ in terms of $r_m$ with
 \bea
{\l \ov 2} & \,=\, & \int_{r_m}^\infty {dr \ov \dr}  \,=\, \int_{r_m}^\infty dr {1 \ov A'} {A' \ov \dr}\,.
\labell{threex16}
 \eea
However, in the present case, it is not possible to use the explicit
root from eq.~\eqref{conserved} for $\dr$ and perform the integral.
Hence we define
 \be
F(r,r_m)\,=\, -\int^\infty_{r} dr {A' \ov \dr} \,, \labell{threex18}
 \ee
and use integration by parts in eq.~\eqref{threex16} to write
 \bea
{\l \ov 2} & \,=\, & \left[ {1 \ov A'}F(r,r_m) \ri]_{r_m}^\infty
+ \int_{r_m}^{\infty} dr {A'' \ov A'^2} F(r,r_m) \nonumber \\
\ & \,=\, & -{F(r_m,r_m) \ov A'(r_m)} + \int_{r_m}^{\infty} dr
{A'' \ov A'^2} F(r,r_m) \,.
\labell{threex17}
 \eea
In eq.~\eqref{threex18}, we have chosen the above limits on integration
because the integrand vanishes at the asymptotic boundary $r=\infty$.
Further, the dependence on $r_m$ in eq.~\eqref{threex18} comes from
$\dr(x,r_m)$. We should remind the reader that in the following
discussion, our independent parameters are the profile $r(x,\l)$ and
$r_m$. Now differentiating eq.~\eqref{threex17}, we find
 \bea
{1\ov 2}{d \l \ov d r_m} & \,=\, & - {1 \ov \dr}
\bigg|_{r_m} - {1 \ov A'(r_m)} {\partial \ov \partial r_m}F(r,r_m)
\bigg|_{r_m} + \int_{r_m}^{\infty} dr {A'' \ov A'^2} {\partial \ov
\partial r_m}F(r,r_m)\,.
 \labell{threex19}
 \eea
The first term above arises since ${\partial F(r,r_m) /
\partial r}=A'(r)/\dr$ but note that this term should be evaluated slightly
away from $r=r_m$ since $\dr=0$ there. This potential divergence will
be canceled below by a contribution which is revealed in the second
term below. Now substituting eqs.~\eqref{threex17} and \eqref{threex19}
into eq.~\eqref{threex15}, we find
 \be
{d {c}_d \ov d r_m } \,=\, {4 \, \pi (d-1) \beta_d \, \l^{d-2} \ov
\lp^{d-1} K_d } \left(I_1 - I_2 \right)\,, \labell{threex20}
 \ee
where
 \bea
I_1 & \,=\, & \int_{r_m}^\infty dr {A'' \ov A'{}^2} {\partial \ov
\partial r_m}F(r,r_m) + A'(r_m)\int_{r_m}^\infty dr {A'' \ov A'^2} F(r,r_m)
 \nonumber \\
\ & \,=\, & -A'(r_m) \int_{r_m}^\infty dr {A'' \ov A'{}^2} \left[
\int^\infty_r d \tilde{r} \, {A'(\tilde{r}) \ov \dtr^2} \left( \dtr
+(d-1) K_d {\partial \dtr \ov \partial K_d } \ri)  \ri]
\labell{threex21}
 \eea
and
 \bea
I_2 & \,=\, & {1 \ov \dr} \bigg|_{r_m} + {1 \ov A'(r_m)} {\partial \ov
\partial r_m}F(r,r_m) \bigg|_{r_m} + F(r_m, r_m) \nonumber \\
\ & \,=\, & {1 \ov \dr} \bigg|_{r_m} - \int^\infty_{r_m} dr \, {A' \ov
\dr^2} \left( \dr +(d-1) K_d {\partial \dr \ov \partial K_d } \ri) \,.
\labell{threex22}
 \eea
In writing these expressions, we have used $\partial \dr / \partial r_m
= (\partial \dr / \partial K_d)(d K_d/ dr_m)$, as well as
eq.~\reef{threex1}. Now considering the local expression for $K_d$ in
eq.~\eqref{conserved}, we calculate $\partial \dr / \partial K_d$
keeping $r$ fixed. Similarly, we can differentiate
eq.~\eqref{conserved} with respect to $x$, which yields an expression
involving $\ddot{r}$. We find that these two quantities are related by
the following:
 \be
\dr \, A' \left( \dr +(d-1) K_d\, {\partial \dr \ov \partial K_d } \ri)
\,=\, \ddot{r}+ 4\lambda L^2\, A'\,A'' \, \dr^2 \, Q
 \labell{threex23}
 \ee
where
 \be
Q \,\equiv\, {\dr^2 + e^{2A} \ov  \dr^2 \, (1-2 \, \lambda L^2 A'{}^2)
+e^{2A} (1 + 4 \, \lambda L^2 \, A'{}^2)  } \,. \labell{threex23x1}
 \ee
We now use eq.~\eqref{threex23} to express eqs.~\eqref{threex21} and
\eqref{threex22} as
  \bea
I_1 & \,=\, & -A'(r_m) \int_{r_m}^{\infty} {dr \ov \dr} {A'' \ov
A'{}^2} - 4\lambda L^2\,A'(r_m)\int_{r_m}^\infty dr \, {A'' \ov A'{}^2}
\left[ \int^\infty_r d \tilde{r} \, {A'(\tilde{r})\,A''(\tilde{r})  \ov
\dtr}\,Q(\tilde{r}) \ri] \,,
 \nonumber \\
I_2 & \,=\, & 4\lambda L^2\,\int_\infty^{r_m} d r \, {A'\,A''  \ov
\dr}\,Q \,,
 \labell{threex25}
  \eea
where we have used $\int dr \, \ddot{r} / \dr^3 = - \int dx \,
\partial/\partial x (1/\dr) = -1/\dr $. Now inserting these expression
\eqref{threex25} into eq.~\eqref{threex20} and then integrating by
parts, we arrive to following result
  \bea
{d{c}_d \ov d r_m } & \,=\, & - {4 \, \pi (d-1) \beta_d \, \l^{d-2} \,
A'(r_m) \ov \lp^{d-1} \,K_d(r_m) }
 \int_{r_m}^\infty d r \, {A'' \ov \dr A'{}^2} \left( 1- 4\lambda L^2\, A'^2 Q \ri)
  \nonumber \\
  & \,=\, & - {4 \, \pi  \beta_d \, \l^{d-2} \,
A'(r_m) \ov \lp^{d-1} \,K_d(r_m) }
 \int_{0}^\l d x \, {1 \ov A'{}^2} \left( T^t{}_t-T^r{}_r\right)
 \,\frac{1- 4\lambda L^2\, A'^2 Q }{1-2\lambda L^2\, A'^2}
  \labell{threex26}
  \eea
Here we used the GB equations of motion, \ie eq.~\reef{null}, to
replace $A''$ by various components of the matter stress tensor. Note
that the final result matches that in eq.~\reef{twox19} when
$\lambda=0$. However, with $\lambda\ne0$, it is clear that the null
energy condition alone (which ensures $T^t{}_t-T^r{}_r\le0$) is
insufficient to enforce a definite sign for $d{c}_d / d r_m$. Rather we
must also be able to make a clear statement about the positivity of the
last factor in the integral, \ie
 \be
\frac{1- 4\lambda L^2\, A'^2 Q }{1-2\lambda L^2\, A'^2}  \,=\, { \dr^2
\, (1-6 \, \lambda L^2 A'{}^2) +e^{2A} \ov (1-2\lambda L^2\, A'^2)(
\dr^2 \, (1-2 \, \lambda L^2 A'{}^2) +e^{2A} (1 + 4 \, \lambda L^2 \,
A'{}^2)) } \,.
 \labell{extra88}
 \ee
As described above, one could use eq.~\eqref{conserved} to express
$\dr^2$ in terms of the conserved charge $K_d$ and the conformal factor
$A$, however, the resulting expression for eq.~\reef{extra88} is
lengthy and unilluminating. In the limit of small $\lambda$ we observe
that eq.~\reef{extra88} becomes
  \be
\frac{1- 4\lambda L^2\, A'^2 Q }{1-2\lambda L^2\, A'^2}  \,\simeq\, 1-2
\, \lambda L^2 A'{}^2+\cdots\,.
 \labell{threex28}
  \ee
Hence it is not clear what simplification one might expect in
eq.~\reef{threex26}. However, this result is still suggestive in that
it is easy to see that the right hand side is positive as long as
$\lambda<0$. Unfortunately, examining the full expression in
eq.~\reef{extra88}, we see that this simple condition does not quite
guarantee that this factor is positive. Thus while we have an
expression for $d{c}_d / d r_m$ in GB gravity, we are not able to make
a simple statement of the conditions that are necessary to ensure that
the c-function flows monotonically along holographic RG flows.

%%%%%%%%%%%%%%%%%%%%%%%%%%%%%%%%%%%%%%%%%%%%%%%%%%%%%%%%%%%%%%%%%%%%%%%%%%%%%

\section{Discussion} \label{discuss}

With eq.~\reef{candid}, we constructed a simple extension to higher
dimensions of the c-function \reef{dim2} considered in
ref.~\cite{casini2} for two-dimensional quantum field theories. As
described in section \ref{ent}, while the entanglement entropy itself
contains a UV divergence, this expression \reef{candid} is finite and,
at conformal fixed points, yields a central charge that characterizes
the underlying conformal field theory, as had been noted previously in
\cite{ent2,Ryu2}. In section \ref{one}, we examined the behaviour of
this c-function in holographic RG flows in which the bulk theory was
described by Einstein gravity. In particular, we were able to show that
the flow of the c-function was monotonic as long as the matter fields
driving the holographic RG flow satisfied the null energy condition. As
reviewed in section \ref{review}, the latter condition was precisely
the constraint that appears in the standard derivation of the
holographic c-theorem \cite{gubser,flow2,cthem}.

We observe that if the bulk geometry is such that it `slightly'
violates the null-energy condition over a `small' radial regime, the
integral in eqs.~\reef{onex15} or \reef{twox19} would remain positive
and hence the flow of our c-function would still be monotonic. That is,
we only need the null energy condition to be satisfied in some averaged
sense. Hence the null-energy condition is a sufficient but not a
necessary condition for the monotonic flow of the c-function
\reef{candid}. Thus there is less sensitivity to the bulk geometry is
the present construction of a holographic c-theorem using the
entanglement entropy than in the original discussions
\cite{gubser,flow2}. It is intriguing that when expressed as an
integral over the boundary direction $x$, eq.~\reef{twox19} weights the
contributions of the bulk stress tensor more or less equally for each
interval $\delta x$ in the strip. However, when the integral is
expressed as an integral over the radial direction, the integrand
includes an extra factor or $1/\dot{r}$, which diverges at the minimum
radius $r_m$ of the holographic surface (but the integral remains
finite). Hence in the holographic flows, the c-function responds
sensitively to changes in the geometry near this radius in the bulk --
a result that can be seen in the explicit flows discussed in section
\ref{examp1}. Hence given the holographic connection between radius in
the bulk and energy scales in the boundary theory, it seems clear that
the flow of this c-function is most sensitive to the lowest energy
modes probed by the entanglement entropy.

Our result for the monotonic flow of c-function in section \ref{one}
refers to the derivative $dc_d/dr_m$, \ie changes in $c_d$ as we change
the minimum energy scale probed by the entanglement entropy. To
describe the flow of $c_d$ completely in terms of the boundary theory,
we would actually like to establish $dc_d/d\ell\le 0$, \ie the
c-function decreases monotonically as we increase the width of the
strip for which the entanglement entropy is evaluated. In this case, we
would be using the width $\ell$ as a proxy for the relevant energy
scale along the RG flow. The desired result can be established in the
present holographic framework, however, as discussed in section
\ref{examp1}, one must be careful to restrict attention to the physical
saddle points in evaluating the entanglement entropy. We showed there
that extremal surfaces can arise for which $d\l/dr_m > 0$ and hence
$dc_d/d\ell>0$. However, these saddle points do not contribute when one
evaluates the holographic entanglement entropy with eq.~\reef{define}
since they are never the minimum area surface. Rather the appearance of
these `unstable' saddle points signals a first order `phase transition'
in the entanglement entropy. As a result, $c_d$ drops discontinuously
at some critical value $\l_t$ of the width of the strip and the
monotonic `flow' of the c-function is preserved.

While we have only illustrated this behaviour with specific examples in
section \ref{one}, it seems clear that the physical entanglement
entropy will never be determined by such saddle points. In particular,
if we are studying a holographic RG flow between two AdS geometries, we
will always find $d\l/dr_m < 0$ when $r_m$ is well into either of these
two asymptotic regions. Hence as argued in section \ref{smooth},  if
extremal surfaces arise for which $d\l/dr_m > 0$, it indicates that
there are a number of competing saddle points in the corresponding
regime. First $\l(r_m)$ is a single-valued function since the conserved
charge \reef{conserved} dictates that there is a unique surface for
each value of $r_m$. Hence if we assume this is a smooth function,
there will always be (at least) three competing saddle points when
$d\l/dr_m > 0$. It then becomes inevitable that there will be a phase
transition in the corresponding regime of $\l$. Further we note that
 \be
\frac{dS\ }{dr_m}=\frac{H^{d-2}}{\l^{d-1}}\,\frac{c_d}{\beta_d}\,
\frac{d\l\ }{dr_m}\,.
 \labell{muck}
 \ee
The first two factors above are positive definite and hence the sign of
$dS/dr_m$ is controlled entirely by $d\l/dr_m$. Given this result, it
is straightforward to argue that the behaviour illustrated in figure
\ref{2ee} is in fact generic. That is, the phase transition goes
between the two branches for which $d\l/dr_m < 0$. Hence we have argued
that given $dc_d/dr_m\ge0$, it also follows that $dc_d/d\l\le0$ for RG
flows dual to Einstein gravity.

One may be concerned that the phase transitions noted above are an
artifact of choosing a background geometry in the bulk which is
unphysical in some way. However, with our analysis in section
\ref{examp1} and appendix \ref{bulk}, we argued that the phase
transitions can arise for holographic backgrounds that have a natural
interpretation as an RG flow in the boundary theory, but also for
backgrounds where the interpretation seems to be more exotic. While
this interpretation was explicitly shown to apply in examples of phase
transitions with the boundary dimension $d\ge3$, constructing further
examples to extend this result to $d=2$ does not seem difficult.
However, we note that these phase transitions are undoubtedly effect of
the large $N$ limit which is implicit in our constructions. However, it
may still be that similar behaviour, \ie rapid transitions in the
entanglement entropy, persists in the RG flows of more conventional
physical systems. In any event, it would be interesting to better
understand these phase transitions in the holographic systems. Such a
transition seems to indicate that quantum correlations in underlying
degrees of freedom change dramatically at some particular energy scale
in the RG flow.

%rm new
It is worthwhile noting that phase transitions in the holographic
entanglement entropy of the kind found here and in \cite{friends} for
RG flows also arise in a variety of other holographic constructions.
The simplest example is to consider the case where the entangling
surface contains two disjoint regions. When the two regions are
relatively close together, saddle point determining the holographic
entanglement entropy will be a single connected bulk surface. However,
as the two regions are moved apart, there is a phase transition to a
second saddle point consisting of two separate bulk surfaces
\cite{head}. A similar phase transition was also found in considering
the holographic entanglement entropy of the strip geometry for a bulk
background corresponding to a confining phase of the boundary theory
\cite{tadashi2,igork}. There is a strong similarity between the results
for these confining theories and the present RG flows since the phase
transition again arises as the width to the strip passes through some
critical value and results in a discontinuous drop in the central
charge $c_d$. Further, in both cases, the phase transition can be
interpreted as being produced by a rapid and drastic restructuring in
the correlations of the low energy degrees of freedom (in comparison to
high energy correlations). Similar results were also found for other
entangling geometries, \ie a circular surface in three-dimensional
confining boundary theory \cite{parnp}. Finally similar phase
transitions in the entanglement entropy have also been found in
holographic superconductors as the temperature is varied
\cite{cliffords} and in the time evolution of holographic quantum
quenches \cite{cliffordx}.

In section \ref{three}, we considered extending our results to
holographic models where the gravitational theory in the bulk is
Gauss-Bonnet gravity \reef{GBAction}. While it is straightforward to
construct an expression \reef{threex26} for $d{c}_d / d r_m$ in GB
gravity, it is evident that the null energy condition is not sufficient
to guarantee a monotonic flow of the c-function. Unfortunately,
eq.~\reef{threex26} does not lend itself to a simple statement of the
conditions that would necessary to ensure that the c-function flows
monotonically along holographic RG flows in these models. Further
insight into this question may be provided by examining explicit
holographic RG flows. In section \ref{examp1}, we assumed that the
holographic backgrounds were solutions of Einstein gravity and hence
the entanglement entropy is determined by eq.~\reef{define}. We could
just as easily assume that the same backgrounds are solutions of GB
gravity and examine the behaviour of the c-function defined by
eq.~\reef{gbstrip}. In particular, it would be interesting to see if
there are violations of the monotonic flow of the c-function in certain
parameter regimes.

Of course, it may not be a surprise that the monotonic flow of the
c-function \reef{candid} is not directly connected to the null energy
condition in GB gravity. As described in section \ref{review}, an
important feature of this theory is that at conformal fixed points, the
dual boundary theory has two distinct central charges, given in
eqs.~\reef{effectc} and \reef{effecta}. Using the null energy
condition, ref.~\cite{cthem} established that the charge denoted $\ads$
would satisfy a c-theorem in these holographic models. However, in
section \ref{ent}, we found that the c-function \reef{candid} actually
corresponds to a nonlinear combination of both central charges. Hence
as we noted at the outset, it was improbable that a simple holographic
c-theorem could be established for GB gravity with the present
construction. Setting holography aside, it is known that for
four-dimensional quantum field theories, there is no possible (linear)
combination of the two central charges, $c=C_T$ and $a=\ads$, that can
satisfy a c-theorem other than $a$ alone \cite{anselmi2}.

Of course, GB gravity only provides an interesting extension of the
usual holographic framework for $d\ge4$. For smaller values of $d$, the
curvature-squared interaction \reef{GBterm} does not contribute to the
gravitational equations of motion because of the topological origin of
this term. It may be of interest to study the behaviour of our
c-function for other gravity theories with higher curvature
interactions for $d=2$ and 3. Interesting families of holographic
models were considered with higher curvature theories of the
three-dimensional gravity in \cite{sinha}. A defining feature of these
theories was that the dual $d=2$ boundary theory should exhibit a
c-theorem. Hence these models may provide an interesting holographic
framework to examine the RG flow of $c_2$. However, the work of
\cite{casini2} indicates that this flow must be monotonic for any
unitary and Lorentz invariant quantum field theory and so confirming
this result here would really be a test that these holographic models
define reasonable boundary theories.

Our construction of the c-function \reef{candid} can be applied quite
generally, \ie outside of the context of holography. While the RG flow
of $c_d$ is not expected to be monotonic in a generic setting, we
observe that the flow can be constrained somewhat following the
analysis of \cite{casini2}. In particular, let us define a new
(dimensionful) function of the following form in arbitrary $d$:
 \be
\hat{c}_d\,\equiv\,\frac{c_d}{\ell^{d-2}}\,=\, {\beta_d \ov H^{d-2}} \l
{d S \ov d \l}\,. \labell{disx1}
 \ee
Of course, for $d=2$, we have $\hat{c}_2=c_2$. In any event, we can
apply directly the analysis of \cite{casini2} to show that
$d\hat{c}_d/d\l\le0$ for any $d$. There, the authors considered unitary
and Lorentz invariant field theories and used sub-additivity of the
entanglement entropy to show that $\hat{c}_2$ is a monotonically
decreasing function as $\l$ increases. In particular, they considered
two specific surfaces $b$ and $c$ with a relative boost, as shown in
figure \ref{boosted}. Further $a$ and $d$ chosen as constant time
surfaces in some frame so that they are Cauchy surfaces whose causal
development corresponds to the intersection and union of the causal
development of the original two boosted surfaces. By construction the
surfaces $a$, $b$ and $c$ just touch the boundary of the causal
development of $d$ on either end. Two important observations are:
First, if we are evaluating the entanglement entropy for these segments
in the Lorentz invariant vacuum state, then it should only depend on
the proper length of the corresponding interval. Second, the
entanglement entropy of any of these surfaces will be the same as for
any other Cauchy surface of the corresponding domains since the time
evolution is assumed to be unitary. Now the authors of \cite{casini2}
show that sub-additivity of the entanglement entropy of these regions
imposes the following relation
 \be
S(b)-S(a) \geq S(\alpha b)-S(\alpha a) \,,
\labell{disx2}
 \ee
where they introduce the ratio of the proper lengths $\alpha=c/a=d/b
\geq 1$. Now if the relative boost is taken to be small, this relation
implies that $\l \, dS/d\l = c_2/3 = \hat{c}_2/3$ must decrease with
increasing $\ell$. Now these elegant arguments can also be applied
without change in considering the entanglement entropy for the strip
geometry in arbitrary dimensions, as long as the boundaries of the
strip are orthogonal to the boost direction.  That is, the segments
$a$, $b$, $c$ and $d$ in figure \ref{boosted} now represent the
orthogonal cross-section of four specific strips. Hence with the
previous arguments, we will recover the relation \eqref{disx2} for
arbitrary dimensions. This implies that with the strip geometry in
arbitrary dimensions, we have $d\hat{c}_d/d\l\le0$ --- the same
observation was made in \cite{tadashi2}. %rm new comment
While we cannot conclude that the original c-function \reef{candid}
must decrease with increasing $\l$, using eq.~\eqref{disx1}, we find an
upper bound on the rate at which ${c}_d$ could increase:
 \be
 \l\,{d {c}_d \ov d \l}\,\le\,(d-2)\,
{c}_d \,. \labell{disx3}
 \ee
This bound applies to any $d$-dimensional quantum field theory, subject
to the provisos of Lorentz invariance, unitarity and sub-additivity of
entanglement entropy.
\FIGURE{
\includegraphics[width=0.7\textwidth]{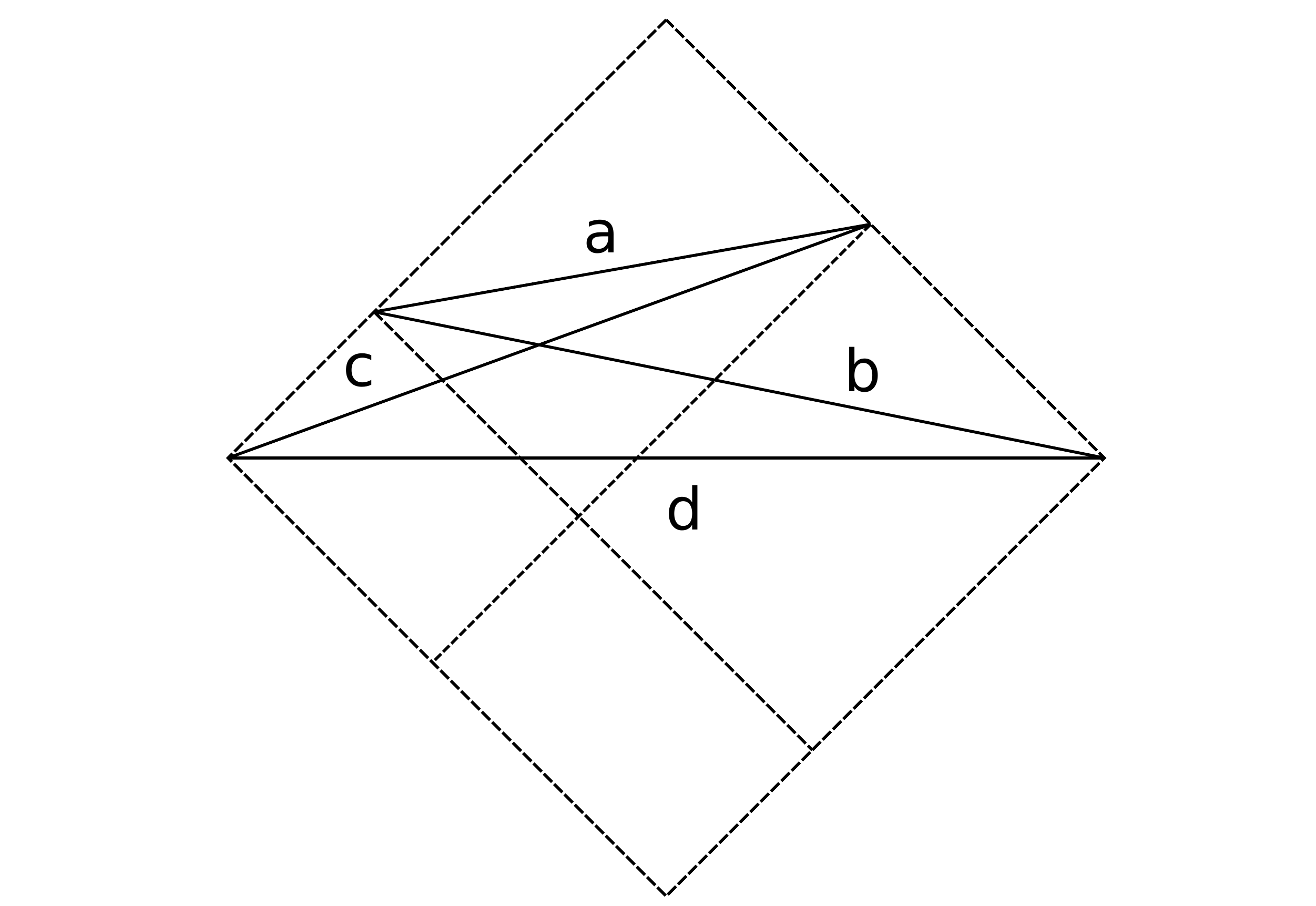}
\caption{Here the boosted surfaces $b$ and $c$ are drawn with time and
space on the vertical and horizontal axes, respectively. The casual
development of the strips $d$ and $a$ contain the union and
intersection of the causal development of $b$ and $c$. The causal
development for each of these surfaces is outlined with dashed lines.}
\label{boosted}}

An important feature of the entanglement entropy \reef{strip} for the
strip geometry is that it contains a single divergent term which is
independent of the width $\ell$. As a result of this simple structure,
the single derivative in eq.~\reef{candid} produces a UV finite or
regulator independent c-function. Now for a general smooth entangling
surface, the structure of the divergent contributions to the
entanglement entropy at a conformal fixed point is more complicated,
\ie
 \be
S\; =\;  \frac{p_{d-2}}{\delta^{d-2}}+ \frac{p_{d-4}}{\delta^{d-4}}
+\ldots
 + \bigg\{
 \begin{array}{l l}
  \frac{p_{1}}{\delta} + p_0 +\mathcal{O}\left(\delta\right)
 &\ \  \textrm{for odd}\ d\,, \\
 \frac{p_{2}}{\delta^2}  + p_0 \log\left(\frac{\l}{\delta}\right)
 +\mathcal{O}(1) &\ \ \textrm{for even}\ d\,,
\end{array}
\labell{entex}
 \ee
where $\delta$ is the short distance cut-off (and $\l$ is some IR
scale). Of course, to produce a dimensionless entanglement entropy, the
coefficients $p_n$ must have dimension length$^n$. At a conformal fixed
point, this dimensionful character is provided by scales arising from
the geometry of the entangling surface and the background spacetime.
For example, the first term yields the celebrated area law with
$p_{d-2}\propto {\cal A}$, where ${\cal A}$ is the area of the
entangling surface. If we move away from a conformal fixed point, there
will be additional dimensionful couplings in underlying theory and the
divergence structure of the entanglement entropy can become
substantially more complicated. For example, in a holographic setting,
the additional complications of relevant operators were illustrated in
\cite{calc}.

Of course, the conjecture by \cite{cthem} proposes that a central
charge, which is identified using the entanglement entropy, should
satisfy a c-theorem \reef{introx1} in higher dimensions . In
particular, the prescription there specifies that the entangling
surface should be a ($d$--2)-sphere of radius $R$ in flat
space.\footnote{The applicability of this geometry to the c-theorem
conjectured in \cite{cthem} follows from the results of
\cite{casini9}.} However the universal central charge was identified
with the dimensionless coefficient $p_0$ which only appears at
subleading order in eq.~\reef{entex} for $d\ge3$. Further since $R$ is
the only scale in the problem, one expects all of the preceding
coefficients are nonvanishing with $p_n=\tilde{p}_n\,R^n$ for some
dimensionless coefficients $\tilde{p}_n$. Hence a single derivative
with respect to the radius of the sphere will not isolate the desired
coefficient in the entanglement entropy. Certainly a more sophisticated
construction would be needed to remove all of the potentially divergent
terms along a general RG flow. In fact, precisely such a construction
was recently proposed in \cite{friends}. It would be interesting to
investigate this new proposal along the lines of the analysis in
sections \ref{one} and \ref{three}. In particular, it would encouraging
if the monotonic flow of the c-function identified with this
construction or a similar variant could be directly related to null
energy condition in holographic RG flows.

\acknowledgments

We would like to thank Alex Buchel, Janet Hung, Luis Lehner, Miguel
Paulos, Subir Sachdev, Amit Sever, Aninda Sinha and Alex Yale for
useful conversations. We also thank Hong Liu for sharing a draft of
\cite{friends} with us before it appeared on the arXiv. Research at
Perimeter Institute is supported by the Government of Canada through
Industry Canada and by the Province of Ontario through the Ministry of
Research \& Innovation. RCM also acknowledges support from an NSERC
Discovery grant and funding from the Canadian Institute for Advanced
Research.

%APPENDIX
%%%%%%%%%%%%%%%%%%%%%%%%%%%%%%%%%%%%%%%%%%%%
\appendix

%%%%%%%%%%%%%%%%%%%%%%%%%%%%%%%%%%%%%%%%%%%%%%%%%%%%%%%%%%%%%%%%%%%%%%%%%%%%%%%%%%%%%%%%%%%%%%%%%%%%

\section{The c-function with Einstein gravity for $d\ge3$} \label{two}

In this appendix, we describe the details of the derivation of
eq.~\eqref{rex1}, which is our starting point for the discussion in
section \ref{one} of the holographic flow of the c-function in higher
dimensions. The analysis here is largely an extension of that given
there for $d=2$ to an arbitrary $d$. For a $d$-dimensional boundary
field theory, the holographic entanglement entropy of the strip is
given by
 \be
S \,=\, {4\pi H^{d-2}\ov \lp^{d-1}} \int_{0}^{(\l-\epsilon)/2} dx \
e^{(d-2)A(r)} \, \sqrt{\dr^2 + e^{2A(r)}}
 \labell{twox1}
 \ee
and the conserved charge \eqref{conserved} reduces to
 \be
K_d \,=\, e^{- d A(r)}  \sqrt{\dr^2 + e^{2A(r)}} \,. \labell{twox3}
 \ee
Further, for a particular extremal surface, $K_d = e^{-(d-1)A(r_m)}$
since at the turning point we have $r(0,\l)=r_m$ and $\dr(0,\l)=0$.

Now we differentiate the entropy \reef{twox1} with respect to the width
$\l$ while holding the radial cut-off $r_c$ fixed. In doing so, we keep
in mind that both the radial profile for the surface and the cut-off in
$x$ are implicitly functions of the width $\l$, \ie $r=r(x,\l)$ and
$\epsilon=\epsilon(\l)$. Hence we find
 \bea
{dS \ov d\l} &\,=\,& {4 \, \pi H^{d-2} \, e^{(d-2) A(r)} \ov \lp^{d-1}
\, \sqrt{\dr^2 + e^{2A(r)}}} \left[ {1\ov 2 }\left(1- {d\epsilon \ov
d\l} \ri) (\dr^2+e^{2A(r)}) + \dr \, {\partial r \ov \partial \l} \ri]
\Bigg|_{x=(\l-\epsilon)/2} \,, \labell{twox5}
 \eea
where $\dr=\partial r(x,\l)/\partial x$. In simplifying ${dS /d\l}$ to
produce the expression above, we have used $\dr(0,\l)=0$ to eliminate
the boundary terms at $x=0$, as well as removing the bulk variation
using the equation of motion which follows from eq.~\reef{twox1},
 \be
\ddot{r} - A'\, \left(d\, \dr^2 +  (d-1)\,e^{2 A} \right)\,=\,0\,.
 \labell{eomd}
 \ee
The cut-off $\epsilon$ is defined by the relation:
$r(x=(\l-\epsilon)/2,\l)=r_c$. We can vary $\l$ in the latter
expression while holding $r_c$ fixed, as in the variation in
eq.~\reef{twox5}, to find
 \be
\left[ {\dr(x,\l) \ov 2} \left( 1 - {d \epsilon \ov d\l}\ri) +
{\partial r(x,\l) \ov \partial \l}\right] \Bigg|_{x={(\l-\epsilon)/ 2}}
\,=\,0\,.
 \labell{onex4a}
 \ee
This result was already presented in eq.~\reef{onex4}. Using this
expression and also eq.~\eqref{twox3}, we can now simplify
eq.~\eqref{twox5} to take the form
 \bea
{dS \ov d\l} &\,=\,&- {4 \, \pi H^{d-2} \ov \lp^{d-1} K_d(\l)} \, {1
\ov \dr} \  {\partial r \ov \partial \l}\, \Bigg|_{x=(\l-\epsilon)/2}
\,. \labell{twox7}
 \eea

Further progress requires that we consider the bulk geometry in the far
UV region, which we assume approaches AdS space asymptotically with
 \be
A(r)\,=\, {r / \tilde{L}}\,. \labell{twox8}
 \ee
Now recall that eq.~\reef{twox3} yields a simple expression for
$dx/dr$, as presented in eq.~\reef{nw1}. With the simple conformal
factor \reef{twox8}, this equation is easily integrated to yield
 \be
x-{\l \ov 2} \,=-\,{\tilde{L} \, e^{-d r/\tL} \ov d \, K_d(\l)} \
{}_2F_{1}\!\left[{1 \ov2 }, {d \ov 2(d-1)}; {3d-2 \ov 2(d-1)};
{e^{-2(d-1)r/\tL} \ov K_d(\l)^2} \ri] \,, \labell{twox9}
 \ee
where we are imposing the boundary condition that $x\to\l/2$ as $r\to
\infty$. We note that the same hypergeometric function appears in
eq.~\reef{byparts2}. This is not a coincidence because in the present
case $A'$ is simply a constant and so we are essentially performing the
same integration here as in eq.~\reef{byparts2}. Further note that this
complicated expression reduces to eq.~\reef{nw2} for $d=2$. Taking the
partial derivative of eq.~\reef{twox9} with respect to $\l$, we find
 \bea
& \ & {\partial r(x,\l) \ov \partial \l} \,=\,-\frac{1}{2} \,e^{dr/\tL}
\, \sqrt{K_d^2 - e^{-2(d-1)r/\tL}}
 \labell{twox11}\\
& \ & + \frac{\tilde{L}}{(d-1)K_d} \, \frac{dK_d}{d\l}  \, \Bigg(
{\sqrt{K_d^2 - e^{-2(d-1)r/\tL}} \ov d\, K_d} \; {}_2F_{1}\left[{1 \ov2
}, {d \ov 2(d-1)}; {3d-2 \ov 2(d-1)}; {e^{-2(d-1)r/\tL} \ov K_d(\l)^2}
\ri]-1 \Bigg)
 \nonumber
 \eea
Now using eq.~\eqref{twox3} with the conformal factor \reef{twox8}, we
find
 \be
\dr \,=\, e^{dr/\tL} \,\sqrt{K_d^2 - e^{-2(d-1)r/\tL}}\,.
 \labell{twox11b}
 \ee
Then we consider the ratio of these two expressions and take the limit
$x \to \l/2$ and $r \to \infty$ to find
 \be
{1 \ov \dr}{\partial r \ov \partial \l} \bigg|_{x=\l} \,=\,-{1 \ov
2}\,. \labell{twox13}
 \ee
Essentially this result indicates that with the limit $r_c\to\infty$,
$d \epsilon/ d\l$ vanishes in eq.~\reef{onex4a}. Now substituting
eq.~\reef{twox13} into eq.~\reef{twox7} yields
 \be
{dS \ov d\l} \,=\, {2 \,\pi H^{d-2} \ov \lp^{d-1}}\,{ 1\ov K_d(\l)}\,.
\labell{twox14}
 \ee
We use this simple form of derivative of entanglement entropy to
produce the expression for the c-function in eq.~\eqref{rex1}.

%%%%%%%%%%%%%%%%%%%%%%%%%%%%%%%%%%%%%%%%%%%%%%%%%%%%%%%%%%%%%%%%%%%%%%%%%%%%%%%%%%%%%%%%%%%%%%%%%%%%

\section{RG flow solutions for GB gravity} \label{gravity}

We are interested in holographic RG flow solutions for the action
\reef{GBAction} in which GB gravity is coupled to a scalar field, \ie
  \be
I=\frac{1}{2\lp^{d-1}}\int d^{d+1}x\sqrt{-g}\left[ R + \frac{\lambda
L^2}{(d-2)(d-3)} \X_4-\frac{1}{2} \left(\del
\phi\right)^2-V(\phi)\right]\,,
 \labell{eqn1}
  \ee
where $\X_4$ is given by eq.~\reef{GBterm}. First, we will examine the
equations of motion for holographic RG flows in some detail. This will
allow us to explicitly verify the result used in eq.~\reef{magic} to
prove the holographic c-theorem for GB gravity \cite{cthem}. Then we
discuss a simple approach to explicitly construct analytic solutions
for GB gravity describing holographic RG flows. This construction is a
simple extension of the `superpotential' approach developed for
Einstein gravity in \cite{Skend,Town1}. This approach was already
extended to GB gravity with four boundary dimensions in \cite{low} and
here we provide the generalization to arbitrary $d\ge4$.

In the action \reef{eqn1} above, the cosmological constant term has
been absorbed into the scalar potential and we assume that $V(\phi)$
has various critical points where the potential energy is negative as
in eq.~\reef{crit2}. As described in section \ref{review}, for each of
these critical points, there is an AdS vacuum solution where the
curvature scale is given by $\tL^2=L^2/\fin$ where $L$ is some
canonical scale appearing in the potential and the curvature squared
interaction while the (dimensionless) constant $\fin$ is given by
eq.~\reef{physical}. To consider solutions describing holographic RG
flows, we begin by writing the scalar and gravitational equations of
motion as
  \bea
\nabla^2\phi -\frac{\delta V}{\delta \phi} &\,=\,& 0\,, \labell{scalar} \\
R_{ab}-\frac12 R g_{ab}+\frac{\lambda L^2}{(d-2)(d-3)}H_{ab} &\,=\,&
T_{ab} \,, \labell{geom}
  \eea
where
  \be
H_{ab}=R_{acde}R_b{}^{cde}-2R_{ac}R_b{}^c-2R_{acbd}R^{cd} +R R_{ab}
-\frac14 \X_4 g_{ab}\,.
 \labell{gbterm2}
  \ee
Further the stress tensor for the scalar field is given by
  \be
T_{ab}=\frac12\del_a\phi\,\del_b\phi-\frac12\,g_{ab} \left(\frac12
\left(\del \phi\right)^2+V(\phi)\right)\,.
 \labell{stress}
  \ee

As in the main text, we consider the following ansatz for the metric:
  \be
ds^2=e^{2A(r)}\,\eta_{ij}\,dx^i dx^j+dr^2\,,
 \labell{eqn2}
  \ee
For nontrivial RG flows, we also include a simple ansatz for the
scalar: $\phi=\phi(r)$. In particular then, this ansatz maintains
Lorentz invariance in the boundary directions. Now with these metric
and scalar ansatze, there are two nontrivial components of the
gravitational equations \reef{geom}:
 \bea
d(d-1)(A'{}^2-\lambda L^2 \,A'{}^4) \; &=&\; 2 \, T_r{}^r \,, \labell{eqn3} \\
2 \, (d-1)(1-2\,\lambda L^2 \, A'{}^2) A'' + d(d-1)(A'{}^2-\lambda L^2
\, A'{}^4) \;&=&\; 2 \, T_t{}^t \,. \labell{eqn4}
 \eea
Again `prime' denotes a derivative with respect to $r$. Using
eq.~\eqref{stress}, we find the following components of stress tensor
 \bea
T_r{}^r&=& \frac14(\phi')^2-\frac12 V(\phi)\,,\labell{rr}\\
T_i{}^j&=& -\,\delta_i{}^j\left(\frac14 (\phi')^2+\frac12
V(\phi)\right)\labell{ij} \,.
 \eea
With the present ansatz, the equation of motion for the bulk scalar
\reef{scalar} becomes
 \be
\phi''+d \: A' \: \phi' - {\delta V \ov \delta \phi} \;=\; 0\,.
\labell{eqn5}
 \ee
Of course, the three equations of motion above are not all independent.
For example, one can derive eq.~\eqref{eqn4} by differentiating
eq.~\eqref{eqn3} and then substituting in eqs.~\eqref{eqn3} and
\eqref{eqn5}.

Taking the difference of eqs.~\eqref{eqn3} and \eqref{eqn4}, we find
 \be
T_t{}^t-T_r{}^r\,=\,(d-1)(1-2\,\lambda L^2 \, A'{}^2) A''\,,
 \labell{null}
 \ee
which is the result used to establish the holographic c-theorem for GB
gravity in eq.~\reef{magic}. Eqs.~\reef{eqn3}, \reef{eqn4} and
\reef{null} are written for GB gravity coupled to a general matter
field Lagrangian and as long as the matter sector satisfies the null
energy condition, the combination of components of $T_{ab}$ in
eq.~\reef{null} are negative (or zero). In the particular case
considered here, \ie the action \reef{eqn1} with a scalar field, we
find $T^t{}_t-T^r{}_r=-(\phi')^2/2$ and so the sign of eq.~\reef{null}
is obvious.

Given eqs.~(\ref{eqn3}--\ref{eqn5}), a simple set of explicit solutions
can be constructed by extending an approach developed for Einstein
gravity in \cite{Skend,Town1}. The key idea is to consider a special
class of scalar potentials that can be defined in terms of a
`superpotential' and then express the solution in terms of this
superpotential. This construction was extended from Einstein gravity to
GB gravity in five dimensions (\ie $d=4$) in \cite{low} and here we
provide the generalization to arbitrary $d\ge4$. First, we write the
scalar potential in terms of a superpotential $W(\phi)$ as follows:
 \be
V(\phi) = 2\,(d-1)^2 \,\left( {\delta W \ov \delta \phi}\ri)^2
\,\left(1-2\,\lambda L^2 W^2 \ri)^2 - d(d-1)W^2 \, (1-\lambda L^2 \,
W^2) \,.
 \labell{messV}
 \ee
With a potential of this form, the equations of motion above can be
re-expressed as first order equations:
 \bea
\phi' &\,=\,& -2\,(d-1) \left(1-2 \, \lambda L^2 \, W^2 \ri) {\delta W
\ov \delta \phi} \,, \nonumber\\
A' &\,=\,& W \,.
 \labell{eqn9x1}
\eea Given these first order equations, we may now solve for the metric
\reef{eqn2} and the scalar profile in quadratures.
%rm new comments
We note that the same equations appear in the construction of domain
wall solutions for `new massive gravity' in higher dimensions
\cite{brazil}.\footnote{We thank Ulysses Camara da Silva for pointing
out this reference.} The action studied there differs from the GB
action \reef{eqn1} by the addition of an action proportional to the
square of the Weyl tensor, \ie
$-\frac{\lambda\,L^2}{(d-2)(d-3)}C_{abcd}C^{abcd}$. This additional
interaction does not influence the equations of motion in the present
setting because the metric ansatz \reef{eqn2} for the holographic RG
flows is conformally flat.

%%%%%%%%%%%%%%%%%%%%%%%%%%%%%%%%%%%%%%%%%%%%%%%%%%%%%%%%%%%%%%%%%%%%%%%%%%%%%%%%%%%%%%%%%%%%%%%%%%%%

\section{Scalar potentials for section 5.2 } \label{bulk}  %\ref{examp1}

In section \ref{smooth}, we analyzed the flow of the entanglement
entropy in various bulk geometries which were defined by specifying an
explicit conformal factor, \eg as in eq.~\reef{nw15}. Here we would
like to show that these profiles can arise as solutions of Einstein
gravity coupled to a scalar field with an appropriate potential. In
particular, we use the `superpotential' approach described at the end
of the last appendix, of course, after setting $\lambda=0$. With the
latter simplification, eqs.~\reef{messV} and \reef{eqn9x1} reduce to
 \bea
V(\phi) &=& 2\,(d-1)^2 \,\left( {\delta W \ov \delta \phi}\ri)^2
 - d(d-1)W^2  \,,
 \labell{messVa}\\
\phi' &\,=\,& -2\,(d-1) {\delta W \ov \delta \phi} \,, \qquad\quad
A' \,=\, W \,.
 \labell{eqn9x1a}
 \eea
Now using these equations, we would like to construct $V(\phi)$ given
$A(r)$.

First from eq.~\reef{eqn9x1a}, we find
 \be
A''={\delta W \ov \delta \phi}\,\phi'=-2(d-1)\,\left({\delta W \ov
\delta \phi}\right)^2 =-{1\ov 2(d-1)}\,(\phi')^2\,.
 \labell{next4}
 \ee
Hence if we let $\phi=\phi_\mt{UV}$ at the critical point in the UV,
\ie $\phi(r\to \infty)\to \phi_\mt{UV}$, then we can write
 \be
\phi(r)\,=\,\phi_\mt{UV}-\int_r^\infty
d\tilde{r}\,\phi'(\tilde{r})\,=\,\phi_\mt{UV}-\int_r^\infty
d\tilde{r}\,\bigg[ -2(d-1)\,A''(\tilde{r}) \bigg]^{1/2}\,.
 \labell{next5}
 \ee
Similarly combining eq.~\reef{messVa} with eqs.~\reef{eqn9x1a} and
\reef{next4}, we may write the value of potential along the flow as
 \be
V(r)=-(d-1)\,A''-d(d-1)\,(A')^2\,.
 \labell{next6}
 \ee
Given these two expressions, one can easily make a parametric plot of
the potential $V$ as a function of $\phi$, at least over the range of
the scalar covered in the holographic flow. Alternatively, if
eq.~\reef{next5} can be integrated analytically and inverted, \ie one
can write $r=r(\phi)$, then eq.~\reef{next6} will yield an analytic
expression for $V(\phi)$. We use both of these approaches to describe
the potential corresponding to the conformal factors presented in
section \ref{smooth}.

Let us begin with the conformal factor given in eq.~\reef{nw15}. In
this case, it is straightforward to integrate eq.~\reef{next5} with the
result being
 \be
\tan\!\left(\frac{\pi}2\,\frac{\phi}{\phi_\mt{UV}}\right)\,=\,e^{r/R}
\qquad {\rm with}\ \ \phi_\mt{UV}= \pi\,\sqrt{2 \gamma (d-1)} \,.
 \labell{eqn9x4}
 \ee
We also note that at the far IR of the holographic flow, $\phi\to0$ as
$r\to-\infty$. Given this expression \reef{eqn9x4} and the conformal
factor \reef{nw15}, it is straightforward to calculate the
superpotential $W(\phi)$ using eq.~\reef{eqn9x1a},
 \be
W(\phi)=\frac1{L}+\frac{\gamma}{R}\,\cos\!
\left(\pi\frac{\phi}{\phi_\mt{UV}}\right)\,,
 \labell{super0}
 \ee
and the potential $V(\phi)$ using eq.~\reef{next6},
 \be
V(\phi)=\frac{(d-1)\gamma}{R^2}\,\sin^2\!
\left(\pi\frac{\phi}{\phi_\mt{UV}}\right)-d(d-1)\left[\, \frac1L
+\frac\gamma{R}\,\cos\!
\left(\pi\frac{\phi}{\phi_\mt{UV}}\right)\,\right]^2\,.
 \labell{eqn9x5}
 \ee
Hence we have produced a analytic result for the scalar potential
necessary to produce the bulk geometry with $A(r)$ as in
eq.~\reef{nw15} as a solution of Einstein gravity. In figure
\ref{pots}, the curve with $\sigma=1$ illustrates the behaviour of the
potential \reef{eqn9x5}. Figure \ref{pots2} also shows the potential
for various values of the parameters.

\FIGURE[!ht]{
\includegraphics[width=0.8\textwidth]{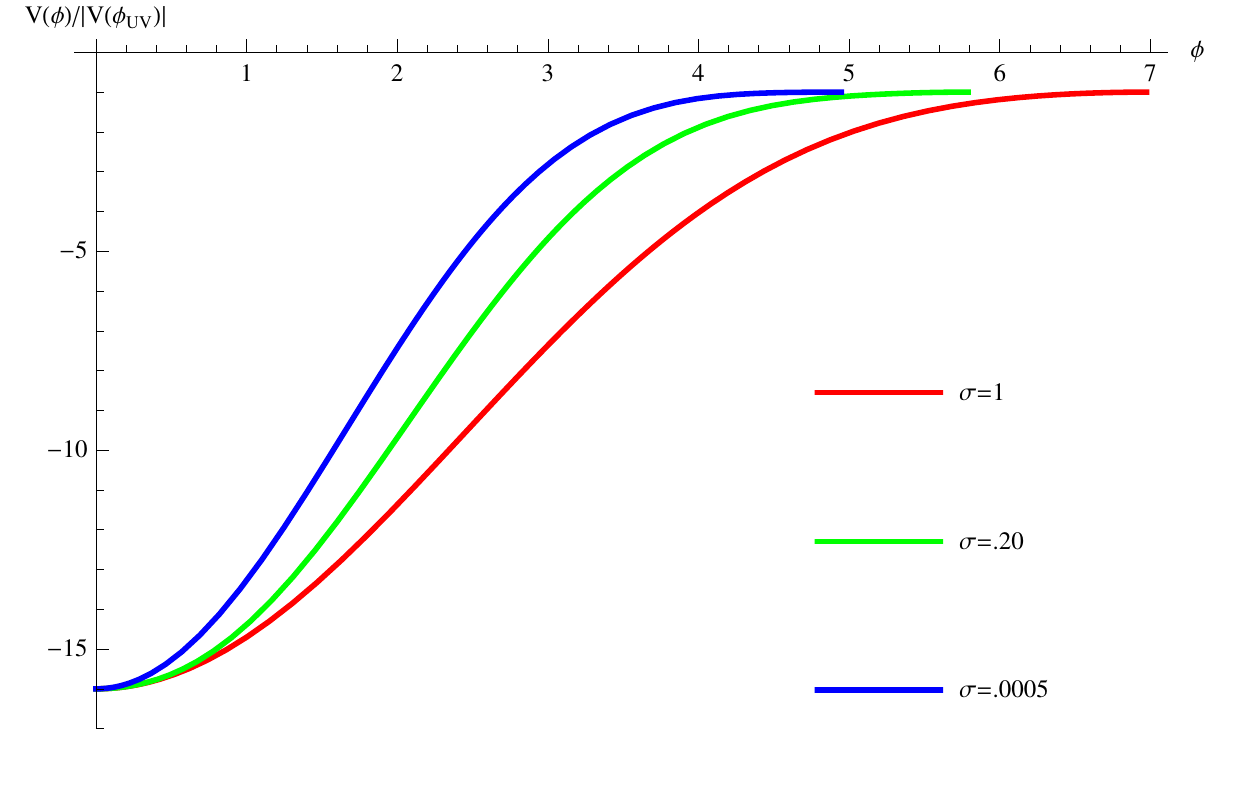}
\caption{A parametric plot of $V(\phi)/|V_\mt{UV}|$ versus $\phi$ for
eq.~\eqref{eqn9x7} with various values of parameter $\sigma$. The other
parameters in the potential are set to $L=.40$, $R=.55$ and
$\gamma=.825$, as well as setting $d=4$.} \label{pots}}
Now let us consider the conformal factor given in eq.~\eqref{nw18}.
Using the same equations as before, we now find
 \bea
{d\phi \ov d r} &\,=\,& {2\sqrt{\gamma(d-1)} \ov R}\,{
\sqrt{1+\sigma\cosh(2r/R)}\ov
\sigma+\cosh(2r/R)}
 \labell{eqn9x7} \\
V(r) &\,=\,& {2\gamma(d-1) \over R^2}\, {1+\sigma \cosh(2r/R) \ov
\left(\sigma +\cosh(2r/R)\right)^2 } - d(d-1) \left(
\frac{1}{L}-{\gamma\ov R}\,\frac{ \sinh\left({2 r}/{R}\right)}{\sigma
+\cosh\left({2 r}/{R}\right)} \right)^2 \,.
 \nonumber
 \eea
In this case, we were not able to analytically integrate $\phi'$.
However, it is straightforward to perform a numerical integration and
produce a parametric plot of the potential $V$ as a function of $\phi$,
as shown in figure \ref{pots}. In this figure, the boundary dimension
was chosen to be $d=4$ and the potential parameters are fixed as
$L=.40$, $R=.55$ and $\gamma=.825$, while $\sigma$ is varied as
indicated on the plot. The plot shows that qualitatively this potential
has the same shape as the analytic potential found for the previous
conformal factor in eq.~\reef{eqn9x5} -- recall that in fact, this new
potential will agree with eq.~\reef{eqn9x5} when $\sigma=1$. However,
the new potential is becoming steeper as the parameter $\sigma$ becomes
smaller. The plots only show the potential over the range relevant for
the holographic RG flow, \ie from $\phi=\phi_\mt{IR}=0$ to
$\phi_\mt{UV}$. However, the full potential would be symmetric about
both $\phi_\mt{IR}$ and $\phi_\mt{UV}$ and so it would be periodic with
a period $\Delta\phi=2(\phi_\mt{UV} - \phi_\mt{IR})$. Further we note
that for the parameters chosen in figure \ref{pots}, the entanglement
entropy undergoes a phase transition for $\sigma=.0005$ but there is no
phase transition for either $\sigma=.2$ or 1.

While we were not able to produce an analytic expression for the full
potential from eq.~\reef{eqn9x7}, it is still possible to consider a
perturbative expansion around the UV critical point, \ie
 \be
V(\dphi)=V_\mt{UV} + \frac12\,m^2_\mt{UV}\, \dphi^2 + \frac1{24}\,
\lambda_\mt{UV}\, \dphi^4+\cdots\,,
 \labell{pertpot1}
 \ee
where $\dphi=\phi-\phi_\mt{UV}$. One can easily verify that there are
no odd powers of $\dphi$ appearing in this expansion. As an example, we
note that the mass parameter above would be calculated as
 \be
m^2_\mt{UV} \,=\, \left. \frac{\delta^2
V}{\delta\phi^2}\right|_{\phi=\phi_\mt{UV}}\,=\, {1 \ov \phi'}\, {d\ov
dr}\left({1\ov \phi'}\,{dV\ov dr} \right) \bigg|_{r\to \infty}\,.
 \labell{eqn9x8}
 \ee
The results for the first few parameter is the expansion
\reef{pertpot1} are:
 \bea
V_\mt{UV} &\,=\,& -{d(d-1)\ov L_{UV}^2}\,,\qquad\qquad  m^2_\mt{UV}
\,=\,{1\ov R^2}-{d \ov L_{UV}R}
 \notag \\
\lambda_\mt{UV}&\,=\,&- {3d\, \sigma^2\, L_{UV}R+d(2-7\sigma^2)\,
L_{IR}R-8(1-2\sigma^2)\,L_{IR}L_{UV} \ov 2 (d-1)\sigma^2\,(L_{UV}-
L_{IR})R^3 } \,,
 \labell{pertpot2}
 \eea
where we are using the expressions in eq.~\reef{nw16} for $L_{UV}$ and
$L_{IR}$. Of course, the value of $V_\mt{UV}$ above corresponds to the
expected value of the cosmological constant in the asymptotic AdS
geometry. We note that the mass parameter $m^2_\mt{UV}$ is independent
of the extra parameter $\sigma$ appearing in the conformal factor
\eqref{nw18}. Hence these first two parameters precisely match those in
the analogous expansion of the analytic potential in eq.~\reef{eqn9x5},
as is readily verified. As shown in eq.~\reef{pertpot2}, the additional
parameter $\sigma$ first makes its appearance the quartic coupling
$\lambda_\mt{UV}$. In this case, one can still verify that
$\lambda_\mt{UV}(\sigma=1)$ matches the analogous quartic coupling
found in expanding the analytic potential \reef{eqn9x5} about
$\phi=\phi_\mt{UV}$. To provide some qualitative insight for this
quartic coupling, we add that typically (\eg for the parameters chosen
in figure \ref{pots}), $\lambda_\mt{UV}$ is negative for $\sigma\sim 1$
but it becomes positive for small values of $\sigma$ (and diverges as
$\sigma\to0$).

At this point, we would like to consider the holographic interpretation
of these geometries in more detail. Recall that the standard
description begins with a discussion in the boundary theory where a UV
critical point is perturbed by some relevant operator $\mathcal{O}$ and
the latter triggers an RG flow to a new critical point in the IR. A
natural assumption in this discussion is that $\mathcal{O}$ is
relevant, \ie that it has a conformal dimension $\Delta<d$. Now in the
gravity description, this operator is dual to the bulk scalar $\phi$
and the conformal dimension is related to the scalar mass by the
standard formula $m_\mt{UV}^2 L^2_{UV}=\Delta(\Delta-d)$ \cite{revue}.
Inverting the latter relation yields two roots for $\Delta$, \ie
 \be
 \Delta_\pm={d\ov2}\pm\sqrt{{d^2\ov4}+m_\mt{UV}^2 L^2_{UV}}\,,
 \labell{stand0}
 \ee
where the standard choice corresponds to $\Delta=\Delta_+$. The scalar
will have two independent solutions asymptotically \cite{revue},
 \be
 \delta\phi \simeq e^{\Delta_-r/L_{UV}}\phi^\mt{(--)} +
 e^{\Delta_+r/L_{UV}}\phi^\mt{(+)}\,,
 \labell{field0}
 \ee
where the coefficient of the more slowly decaying solution
$\phi^\mt{(--)}$ corresponds to the coupling for the dual operator
while $\phi^\mt{(+)}$ is proportional to the expectation value
$\langle\mathcal{O} \rangle$.

\FIGURE[!ht]{
\includegraphics[width=0.9\textwidth]{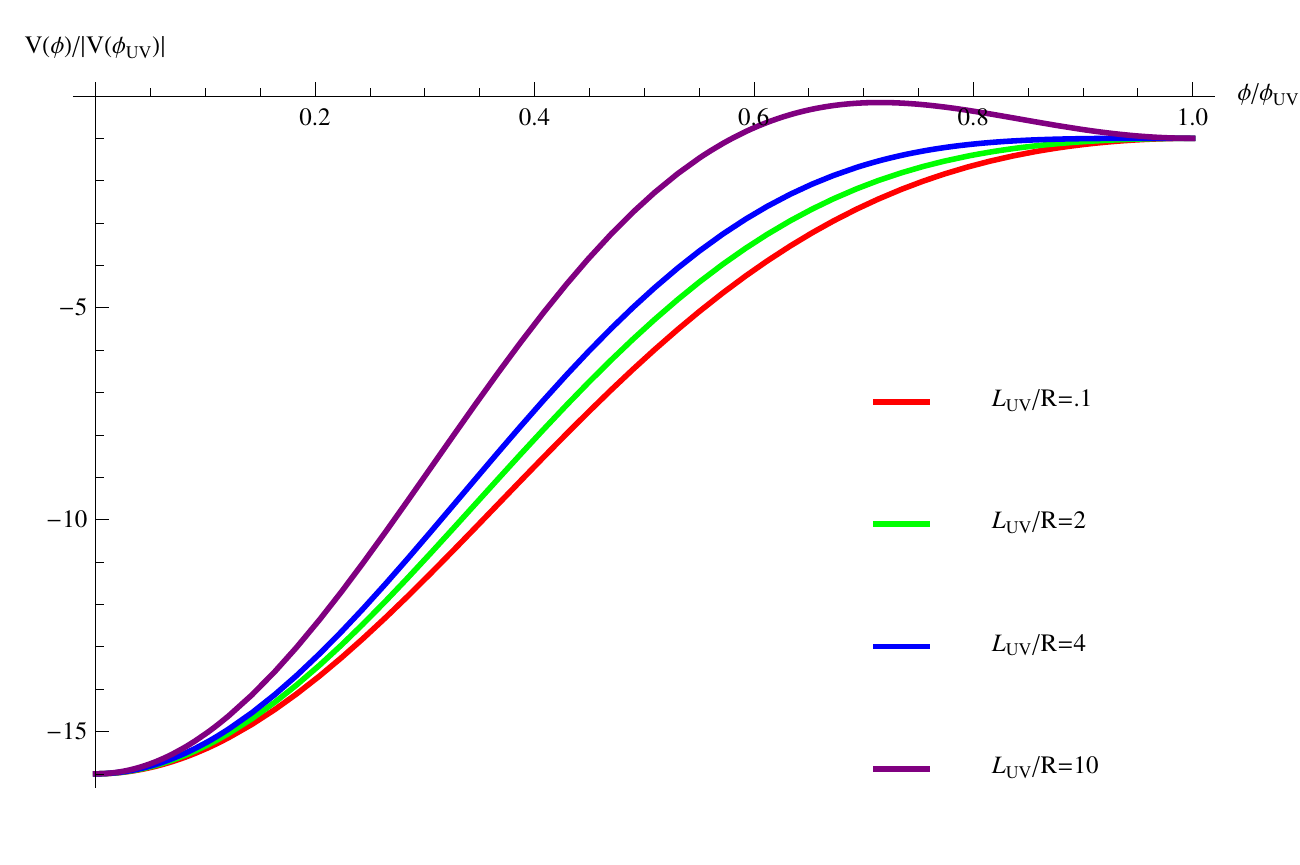}
\caption{Plot of $V(\phi)/|V_\mt{UV}|$ versus $\phi/\phi_\mt{UV}$ for
eq.~\reef{eqn9x5} with various values of parameter $L_{UV}/R$. The
parameters in the potential are chosen to fix $L_{UV}=1$ and
$L_{IR}=.25$ and we set $d=4$.} \label{pots2}}
Now turning to the scalar mass in eq.~\reef{pertpot2}, we observe that
the result can be written as $m_\mt{UV}^2 L^2_{UV}=\alpha(\alpha-d)$
where $\alpha=L_{UV}/R$. Further by combining eq.~\reef{nw17a} and
$\phi'\propto \sqrt{-A''}$ from eq.~\reef{next4}, we see that $\alpha$
is precisely the exponent controlling the asymptotic decay of
$\delta\phi$. Now there are four regimes of $\alpha$ which we consider
separately: (a) $0<\alpha\le d/2$, (b) $d/2< \alpha<(d+2)/2$, (c)
$(d+2)/2< \alpha<d$ and (c) $d\le\alpha$. In the interval (a), the
holographic interpretation of the flow geometry precisely matches that
described above. That is, $\alpha=\Delta_-$ with
$\Delta_+=d-\Delta_-<d$. Hence the dual operator is relevant and
leading contribution in the asymptotic decay of $\delta\phi$ reflects
the fact that corresponding coupling is nonvanishing in the boundary
theory. In the intervals (b) and (c), we have instead
$\alpha=\Delta_+<d$ and hence the interpretation is that $\mathcal{O}$
is relevant but the boundary coupling for this operator vanishes.
Rather holographic interpretation of the bulk solution is that
$\mathcal{O}$ has a nonvanishing expectation value in the UV, which
then triggers the RG flow to an new IR fixed point. Of course, this
appears to be a somewhat unconventional description of RG
flows.\footnote{Similar flows arising in a supergravity construction
were studied in \cite{ictp}. However, the flows described there would
all fall into the interval (b).} We have distinguished the interval (b)
with ${d\ov2}< \alpha<{d\ov2}+1$ because in the regime, one can imagine
the standard interpretation of the bulk solution still holds for the
`alternate quantization' of the holographic theory \cite{igor9}. In the
latter approach, the roles of $\phi^\mt{(--)}$ and $\phi^\mt{(+)}$ are
reversed and the dimension of the boundary operator is given by
$\Delta=\Delta_-$. Note that in these first three intervals (a), (b)
and (c), we have $m^2_\mt{UV}\le0$ and $m^2_\mt{UV}$ reaches its
minimum value at $\alpha=d/2$, where it coincides precisely with the
Breitenlohner-Freedman bound \cite{BF}, \ie
$m^2_\mt{UV}L^2_{UV}=-d^2/4$. Finally in the interval (d), we have
$\alpha=\Delta_+\ge d$ and hence in this regime, $\mathcal{O}$ is no
longer a relevant operator. Rather we have an irrelevant (or marginal)
operator again with a nonvanishing expectation value which triggers the
flow to a new fixed point in the IR. Note that with $\alpha>d$,
$m_\mt{UV}^2>0$ and the scalar potential must have a third extremum
$\phi_3$ between the UV and IR critical points where
$V(\phi_3)>V_\mt{UV}$.\footnote{In fact, for $L_{UV}/R$ sufficiently
large, one finds $V(\phi_3)>0$.} This extra extremum is distinguished
from the UV and IR critical points since by construction, the latter
are critical points of the superpotential, whereas the new critical
point satisfies $\delta V/\delta\phi|_{\phi=\phi_3}=0$ but $\delta
W/\delta\phi|_{\phi=\phi_3} \ne0$. The changing structure of the
potential as $L_{UV}/R$ varies through these different regimes is
illustrated in figure \ref{pots2}. As already noted, the interpretation
of the RG flows in cases (c) and (d) in the boundary theory seems
somewhat exotic. Hence one must worry that the underlying holographic
model for these constructions is unphysical in some way. For example,
it could be that for a consistent boundary CFT, once the dimension
$\Delta$ of the operator is fixed, the quartic coupling in
eq.~\reef{pertpot2} must be constrained in some way, along the lines of
various bounds found in \cite{EtasGB} or \cite{boudn}. Alternatively,
it could be that these background solutions are simply unstable in the
corresponding parameter regime.\footnote{We thank Alex Buchel for
discussions of these issues.} In any event, in the main text, we focus
our attention on the models in the intervals (a) and (b) for which the
holographic RG flows have a conventional interpretation -- see the
discussion around eq.~\reef{bfbound}.

To close this section, we note that one can also consider a
perturbative expansion of the scalar potential as in
eq.~\reef{pertpot1} but about the IR critical point, \ie
 \be
V(\dtphi)=V_\mt{IR} + \frac12\,m^2_\mt{IR}\, \dtphi^{\,2} +
\frac1{24}\, \lambda_\mt{IR}\, \dtphi^{\,4}+\cdots\,,
 \labell{pertpot3}
 \ee
where $\dtphi=\phi-\phi_\mt{IR}$ with
$\phi_\mt{IR}\equiv\phi(r\to-\infty)$. As before, there are no odd
powers of $\dtphi$ appearing in eq.~\reef{pertpot3}. The first few
parameters in this IR expansion are:
 \bea
V_\mt{IR} &\,=\,& -{d(d-1)\ov L_{IR}^2}\,,\qquad\qquad  m^2_\mt{IR}
\,=\,{1\ov R^2}+{d \ov L_{IR} R}
 \notag \\
\lambda_\mt{IR}&\,=\,&- {-d\, (2-7 \sigma^2)\, L_{UV}R-3d\,\sigma^2\,
L_{IR}R-8(1-2\sigma^2)\,L_{IR}L_{UV} \ov 2 (d-1)\sigma^2\,(L_{UV}-
L_{IR})R^3 } \,.
 \labell{pertpot4}
 \eea
As in the UV expansion, we see that the parameter $\sigma$ first
appears here in the quartic coupling $\lambda_\mt{IR}$. Hence one can
easily verify that $V_\mt{IR}$ and $m^2_\mt{IR}$ precisely match the
corresponding couplings in the IR expansion of the analytic potential
in eq.~\reef{eqn9x5}, while $\lambda_\mt{IR}$ matches the quartic
coupling in this expansion when $\sigma=1$. Note that $m^2_\mt{IR}>0$
and so the dual operator is always irrelevant at the IR critical point.
The behaviour of the quartic coupling is qualitatively the same as
described above for $\lambda_\mt{IR}$, \ie it is typically negative for
$\sigma\sim1$ and positive for $\sigma\sim 0$.

%%%%%%%%%%%%%%%%%%%%%%%%%%%%%%%%%%%%%%%%%%%%%%%%%%%%%%%%%%%%%%%%%%%%%%%%%%%%%

\end{document}